# User Selection for Multi-user MIMO Downlink with Zero-Forcing Beamforming

Shengchun Huang, *Student Member, IEEE*, Hao Yin, Jiangxing Wu,
Victor C. M. Leung, *Fellow, IEEE*

**Abstract**

In this paper, we propose a greedy user selection with swap (GUSS) algorithm based on zero-forcing beamforming (ZFBF) for the multi-user multiple-input multiple-output (MIMO) downlink channels. Since existing user selection algorithms, such as the zero-forcing with selection (ZFS), have 'redundant user' and 'local optimum' flaws that compromise the achieved sum rate, GUSS adds 'delete' and 'swap' operations to the user selection procedure of ZFS to improve the performance by eliminating 'redundant user' and escaping from 'local optimum', respectively. In addition, an effective channel vector based effective-channel-gain updating scheme is presented to reduce the complexity of GUSS. With the help of this updating scheme, GUSS has the same order of complexity of ZFS with only a linear increment. Simulation results indicate that on average GUSS achieves 99.3 percent of the sum rate upper bound that is achieved by exhaustive search, over the range of transmit signal-to-noise ratios considered with only three to six times the complexity of ZFS.

S. Huang and H. Yin are with the School of Electronic Science and Engineering, National University of Defense Technology, Changsha, Hunan 410073, P. R. China (e-mail: huangsc@nudt.edu.cn, yinhao1@263.net).

J. Wu is with the National Digital Switching System Engineering & Technology Research Center, Zhengzhou, Henan, P. R. China (e-mail: wujiangxing@126.com).

V. C. M. Leung is with the Department of Electrical and Computer Engineering, University of British Columbia, Vancouver, BC V6T 1Z4, Canada (e-mail: vleung@ece.ubc.ca).

## I. Introduction

Multi-user multiple-input multiple-output (MIMO) communication, where a multi-antenna base station (BS) communicates with multiple users simultaneously, is a key technology to provide high throughput for future wireless communication systems [1]. In this scenario, the BS is usually equipped with more antennas than that supporting single user communication, due to the equipment size, power supply and computation capacity factors. Consequently, it can transmit different data streams to multiple users simultaneously in the downlink to exploit the extra spatial degrees of freedom. A fundamental problem arising in this scenario is how the BS should choose a subset of users for transmissions in order to maximize the total throughput [2]–[5].

The choice of the best user subset $S^{best}$ depends on the precoding method adopted in the BS. Even though dirty paper coding (DPC) [6] is the optimal scheme in the sense that DPC achieves the capacity of MIMO broadcast channel [7]–[10], it is difficult to implement it in practical systems due to its high computational complexity. We consider in this paper a practical low complexity scheme termed as zero-forcing beamforming (ZFBF) [11]–[15], which completely removes the interference by inverting the channel matrix at the transmitter. The number of users that BS can communicate with simultaneously is equal to or less than the number of antennas in BS when the ZFBF precoding is adopted.

Determining $S^{best}$ for the multi-user MIMO downlink with ZFBF requires a brute-force exhaustive search over all possible user sets, and the complexity of an exhaustive search is prohibitive when the number of users is large. Thus, several suboptimal greedy user selection algorithms have been designed in the past. Generally, these algorithms fall into two categories: capacity-based algorithm and Frobenius norm-based algorithm. The capacity-based algorithm, represented by the zero forcing with selection (ZFS) algorithm proposed by Dimic *et al.* [2], chooses users greedily based on the accurate sum rate variation. It chooses the first user with the highest channel capacity and then finds the next user that provides the maximum sum rate from the remaining unselected users. Based on ZFS, Wang *et al.* proposed a sequential water-filling user selection (SWF) algorithm to improve the achieved sum rate performance by eliminating users allocated with zero transmit power after ZFS user selection [5]. The Frobenius norm-based algorithm, represented by the semi-orthogonal user selection (SUS) algorithm proposed by Yoo *et al.* [3], chooses users greedily based on the approximate sum rate variations with respect to channel norm related parameters. SUS adds the new user with the largest effective channel norm nearly orthogonal to the selected users in each iteration. Along this line, Akhlaghi *et al.* proposed a greedy algorithm based on maximizing the determinant of the composite channel matrix [16], and Jin *et al.* proposed a capacity-based algorithm maximizing the product of diagonal elements of the upper-triangular matrix R after performing QR factorization to the channel matrix [17]. The Frobenius norm-based algorithms have lower complexity by eliminating the calculation of sum rate, but pay a price in sum rate performance by not guaranteeing a positive sum rate increment in the user selection process.

Two main flaws exist in previous greedy search user selection algorithms:

- Redundant users exist in selected user set;
- The selected user set might be trapped in a local optimum.

A ‘redundant user’ is defined as a user who can be deleted from the selected user set to yield an increase in the sum rate. Existence of redundant users is an inherent flaw of greedy incremental algorithms since the accumulated user selection procedure would make some former selected users undesirable. This phenomenon has been identified in [2] and [5] that redundant users exist when some users are assigned with zero transmit power after waterfilling power allocation, and solved by deleting the user with zero transmit power. However, as we will prove in Sections III, [2] and [5] were incorrect in both identifying and handling the redundant users, which may exist even though all users are allocated with positive power and it may not achieve the maximum sum rate increment by deleting users with zero power.

Since user selection is a combinatorial optimization problem, the achieved user set of previous algorithms may be trapped in a local optimum where the sum rate cannot be increased by adding a new user or deleting a selected user. However, the sum rate can be increased by swapping users between the selected user set and the candidate users. After leaving the local optimum by a ‘swap’ operation, the ‘add’ and ‘delete’ operation can be utilized further to increase the sum rate.

The main contributions of this paper are as follows:

1. We propose a user selection algorithm with high throughput and low complexity

In this paper, we propose a new user selection algorithm, named greedy user selection with swap (GUSS), which introduces ‘add’, ‘delete’ and ‘swap’ operations in the user selection procedure to increase the sum rate. GUSS eliminates all the redundant users through the ‘delete’ operation and escapes from local optima through the ‘swap’ operation.

2. We present an efficient effective-channel-gain updating strategy to reduce the complexity of GUSS

To avoid expensive matrix inversion involved in updating the sum rate, we design an efficient effective-channel-gain updating method that replaces matrix inversion with less expensive vector-vector multiplication. Previous complexity reduction methods, such as those proposed for ZFS and SWF, are only suitable for incremental user set update while deleting or swapping users cannot be supported. Our method provides the same low complexity for ‘add’, ‘delete’ and ‘swap’ operations.

The remainder of this paper is organized as follows. In Section II, we describe the system model and formulate the user selection problem in multi-user MIMO downlink with ZFBF. The two flaws in existing user selection algorithms are explored in Sections III. In Section IV, the effective-channel-gain updating method for ‘add’, ‘delete’ and ‘swap’ operation is derived. In Sections V, the GUSS algorithm is presented. The sum rate performance and complexity of GUSS are evaluated and compared with previous user selection algorithms in Sections VI. Section VII concludes the paper.

## II. System Model and Problem Formulation

### *A. Notation*

We use uppercase boldface letters for matrices and lowercase boldface for vectors. $E\{\cdot\}$ stands for the expectation operator, $\mathbf{H}^*$ ($\mathbf{h}^*$) stands for the conjugate transpose of a matrix $\mathbf{H}$ (vector $\mathbf{h}$), and $|S|$ denotes the cardinality of a user set $S$. $||\mathbf{h}||$ denotes the Euclidean vector norm that $||\mathbf{h}|| = \sqrt{\mathbf{h}\mathbf{h}^*}$ when $\mathbf{h}$ is a row vector. $\mathbf{H}^\dagger$ denotes the Moore-Penrose pseudo-inverse $\mathbf{H}^\dagger = \mathbf{H}^*(\mathbf{H}\mathbf{H}^*)^{-1}$. $S_1 \setminus S_2$ denotes set difference that deletes the elements of $S_2$ from $S_1$.

### *B. System Model*

Consider a single cell MIMO downlink channel with $M$ transmit antennas at the base station (BS) serving $K$ single antenna users. Assume a quasi-static flat-fading channel between the BS and the users, and $h_{k,m}$ represents the complex channel gain from transmit antenna $m$ to user $k$. Thus, the received signal $y_k$ at user $k$ is determined by

$$y_k = \mathbf{h}_k \mathbf{x} + n_k \tag{1}$$

for $k = 1, \cdots, K$, where $\mathbf{x} \in \mathbb{C}^{M \times 1}$ is the transmitted signal vector, $\mathbf{h}_k = [h_{k,1} \ \cdots \ h_{k,M}] \in \mathbb{C}^{1 \times M}$ is the channel vector of user $k$, and $n_k$ is the white Gaussian noise with zero mean and unit variance. $\mathbf{H} = [\mathbf{h}_1^*, \cdots, \mathbf{h}_K^*]^* \in \mathbb{C}^{K \times M}$ is the channel matrix of all users, the entries of $\mathbf{H}$ are modeled as a set of i.i.d. zero-mean circularly symmetric complex Gaussian random variables and the BS is assumed to have full knowledge of $\mathbf{H}$. The power constraint for the transmitted signal is $E\{\mathbf{x}^*\mathbf{x}\} \leq P$. Since the noise has unit variance, $P$ also means total transmit signal-to-noise ratio (SNR) [7].

The BS supports up to $M$ users simultaneously when using linear beamforming transmission. Denote the index set of served users as $S = \{\pi(1), \cdots, \pi(k)\}$, $k = |S| \leq M$ and $S \subset \{1, \cdots, K\}$. The transmit signal vector $\mathbf{x}$ is a linear combination of all selected users' data streams, constructed as

$$\mathbf{x} = \sum_{i \in S} \mathbf{w}_i \sqrt{p_i} s_i \, , \tag{2}$$

where $\mathbf{w}_i \in \mathbb{C}^{M \times 1}$ is the beamforming weight vector, $p_i$ is the transmit power scaling factor and $s_i$ is the information symbol of user $i$. We can rewrite (1) as

$$y_k = (\mathbf{h}_k \mathbf{w}_k \sqrt{p_k}) s_k + \sum_{i \in S, i \neq k} (\mathbf{h}_k \mathbf{w}_i \sqrt{p_i}) s_i + n_k \, . \tag{3}$$

Finding the optimal beamforming weight vector $\mathbf{w}_i$ is a difficult non-convex optimization problem [3]. We utilize ZFBF, which is easy to implement and has comparable performance with DPC [3], to determining the beamforming weight vectors in this paper.

### C. Zero-Forcing Beamforming

ZFBF inverts the channel matrix at the transmitter in order to create orthogonal channels between the BS and users. ZFBF completely removes the interference among different users at the BS, i.e.,

$$\mathbf{h}_j \mathbf{w}_i = \delta_{i,j}, \quad i, j \in S \, . \tag{4}$$

Therefore, $\mathbf{w}_i^*$ must lies in the orthogonal complement of the subspace $V_i = span\{\mathbf{h}_j | j \in S, j \neq i\}$, denoted it as $V_i^{\perp}$, where $V_i$ is spanned by the channels of all the other selected users [18]. The orthogonal projector matrix on $V_i^{\perp}$ is

$$\mathbf{P}_i^{\perp} = \mathbf{I}_M - \mathbf{H}_{S\backslash\{i\}}^* (\mathbf{H}_{S\backslash\{i\}} \mathbf{H}_{S\backslash\{i\}}^*)^{-1} \mathbf{H}_{S\backslash\{i\}} \ , \tag{5}$$

where $\mathbf{I}_M$ is the $M \times M$ identity matrix, and $\mathbf{H}_{S\backslash\{i\}}$ is the row-reduced channel matrix of all the selected users except user $i$. Suppose $\pi(l) = i$, we have

$$\mathbf{H}_{S\backslash\{i\}} = [\mathbf{h}_{\pi(1)}^*, \cdots, \mathbf{h}_{\pi(l-1)}^*, \mathbf{h}_{\pi(l+1)}^*, \cdots, \mathbf{h}_{\pi(k)}^*]^* \ . \tag{6}$$

Since ZFBF is a linear precoder that maximizes the output SNR subject to the constraint that does not interfere with all other streams [19], according to the orthogonal condition (4) we have [7]

$$\mathbf{w}_i = \left( \frac{\mathbf{h}_i \mathbf{P}_i^{\perp}}{\mathbf{h}_i \mathbf{P}_i^{\perp} \mathbf{h}_i^*} \right)^* = \frac{\mathbf{P}_i^{\perp} \mathbf{h}_i^*}{\mathbf{h}_i \mathbf{P}_i^{\perp} \mathbf{h}_i^*} \ . \tag{7}$$

Define

$$\boldsymbol{\nu}_i = \mathbf{h}_i \mathbf{P}_i^{\perp} \ . \tag{8}$$

The $\boldsymbol{\nu}_i$ can be interpreted as the *effective channel vector* (ECV) of user $i$ . The ECV $\boldsymbol{\nu}_i$ is the component of $\mathbf{h}_i$ orthogonal to $V_i$ and the module square of $\boldsymbol{\nu}_i$ equals to effective-channel-gain $\lambda_i$ as we will prove later in (11). Fig. 1 shows an example of ECV for user 1 and 2 when the selected user set $S = \{1, 2\}$. According the definition in (8), we have $\boldsymbol{\nu}_i \mathbf{h}_j^* = 0$ for all $i \neq j$ , $i, j \in S$ and $\boldsymbol{\nu}_i$ changes with selected user set $S$ that its module decreases when $S$ been added with more users. The beamforming weight vector $\mathbf{w}_i$ can be rewritten as

$$\mathbf{w}_i = \frac{\mathbf{P}_i^{\perp} \mathbf{h}_i^*}{\mathbf{h}_i \mathbf{P}_i^{\perp} \mathbf{h}_i^*} = \frac{\boldsymbol{\nu}_i^*}{\|\boldsymbol{\nu}_i\|^2} \ . \tag{9}$$

The received signal for user $i$ is then given by $y_i = \sqrt{p_i} s_i + n_i$, and the maximum achievable ZFBF sum rate for the user set $S$ is the sum of individual rates

$$R(S) = \max_{p_i : \sum_{i \in S} \lambda_i^{-1} p_i \leq P} \sum_{i \in S} \log(1 + p_i) \ , \tag{10}$$

where

$$\lambda_i = \frac{1}{\|\mathbf{w}_i\|^2} = \|\boldsymbol{\nu}_i\|^2 \tag{11}$$

is the effective-channel-gain of user $i$ [3], $\lambda_i^{-1} p_i$ is the transmit power allocated to user $i$, and $p_i$ is the received SNR of user $i$. By using Lagrangian method, the optimal $p_i$ in (10) is found by waterfilling power allocation

$$p_i = (\mu\lambda_i - 1)^+ = \left(\mu \left\|\boldsymbol{\nu}_i\right\|^2 - 1\right)^+ \ , \tag{12}$$

where $(x)^+$ denotes $\max\{x, 0\}$, and $\mu$ is the water level satisfing

$$\sum_{i \in S} \left(\mu - \|\boldsymbol{\nu}_i\|^{-2}\right)^+ = P \ . \tag{13}$$

Note that there is another simple explicit formula for the beamforming weight vectors: $\mathbf{w}_{\pi(i)}$ is the $i$-th column of the Moore-Penrose pseudo-inverse $\mathbf{H}_S^\dagger$ of the channel matrix $\mathbf{H}_S$, defined by $\mathbf{H}^\dagger = \mathbf{H}_S^*(\mathbf{H}_S\mathbf{H}_S^*)^{-1}$, i.e., $\mathbf{H}_S^\dagger = [\mathbf{w}_{\pi(1)}, \cdots, \mathbf{w}_{\pi(k)}]$. According to (9) and (11), we have

$$\mathbf{H}_S^\dagger = [\frac{\boldsymbol{\nu}_{\pi(1)}^*}{\lambda_{\pi(1)}}, \cdots, \frac{\boldsymbol{\nu}_{\pi(k)}^*}{\lambda_{\pi(k)}}] \ . \tag{14}$$

### *D. Sum rate maximization with user selection*

The sum rate (10) of ZFBF can be further optimized with respect to the selected user set $S$. Thus, the user selection problem can be formulated as

$$\begin{aligned} & maximize && R(S) \\ & subject\ to && S \subset \{1, \cdots, K\} \ . \end{aligned} \tag{15}$$

This is a fundamental question in multi-user MIMO communication, but determining the optimal $S^{best}$ in (15) requires an exhaustive search over all possible user sets. The size of the search space is $\sum_{i=1}^{M} \frac{K!}{i!(K-i)!}$, which increases exponentially with $M$. It is prohibitive for practical implementation. Many suboptimal user selection strategies had been proposed to approach the upper bound set by exhaustive search. A major class of ZFBF user selection method is the incremental heuristic search method [2]–[5], [16], [17], represented by the ZFS algorithm proposed in [2].

## III. Flaws in Previous Greedy User Selection Algorithms

In this section, we study the problems in a typical greedy user selection algorithm represented by ZFS. ZFS is initialized with the user with the maximum channel norm. In each iteration one user is added to the selected user set such that the sum rate increment is maximized. The ‘add’ operation continues until no positive sum rate increment can be achieved. The essential recursive user set updating step of ZFS is

$$\begin{aligned} \pi(n) &= \max_{u \in U \setminus S_{n-1}} R(S_{n-1} \cup \{u\}) \\ S_n &= S_{n-1} \cup \{\pi(n)\} \ , \end{aligned} \quad (16)$$

where $U = \{1, \cdots, K\}$ is the index set of all users, $\pi(n)$ is the index of selected users in the $n$-th step and $S_n$ is the updated index set after adding the selected user $\pi(n)$. Suppose the output of ZFS user selection procedure is $S_{_{ZFS}}$.

Let $U_n$ denote the index set that maximizes the sum rate among all user sets with cardinality $n$, i.e., $U_n = \arg\max_{S \subset U, |S|=n} R(S)$. The essential idea behind (16) is trying to obtain $U_n$ based on $U_{n-1}$ by adding a new user. However, since $U_n$ may not be the superset of $U_{n-1}$, i.e., $U_{n-1} \not\subset U_n$, as we will see later in Fig. 2, the $S_n$ selected by ZFS may not be identical to $U_n$ except when $n = 1$. Furthermore, the $S_{_{ZFS}}$ may not be $S^{best}$ because the optimum $S^{best}$ in (15) achieved by exhaustive search should satisfy $S^{best} = \arg\max_{1 \leq n \leq M} R(U_n)$. The typical flaws in the output of ZFS $S_{_{ZFS}}$ include following two aspects.

### *A. Redundant user*

Because the greedy incremental user selection considers only the influence of selected users, but not including the influence of user yet to be selected, a previously selected user might become a redundant user when new users are added. This phenomenon has been partially discovered in [2] and [5], where they found the existence of redundant users when some users $i \in S$ been assigned with zero transmit power, i.e., $p_i = 0$, after waterfilling power allocation. The redundant user situation is handled by deleting users with $p_i = 0$, and the obtained result is viewed as 'optimal beamforming vector' in [5]. However, as we will prove in the following, there are more to be discovered in both identifying and handling the redundant users.

1. Redundant users might exist even if $p_i > 0$ for each selected user

The condition $p_i = 0$ is sufficient but not necessary for the user $i \in S$ to be redundant. Its sufficiency had been proved in both [2] and [5] that the sum rate will increase after deleting users with $p_i = 0$. It is, however, not a necessary condition, which will be demonstrated in the following.

Let

$$H = \begin{bmatrix} 1 & 0.65 & 0 \\ 0.46 & 1 & 0.46 \\ 0 & 0.65 & 1 \end{bmatrix}, \tag{17}$$

be a channel matrix instant between a three-antenna BS and three single-antenna users. The sum rates for user sets $\{2\}, \{1,2\}, \{1,3\}$ and $\{1,2,3\}$ under different sum transmit SNR $P$ are shown in Fig. 2.

The user set found by exhaustive search, $S^{best}$, varies with transmit SNR $P$ that $S^{best} = \{1,3\}$ for $0\,dB \leq P \leq 34.85\,dB$ and $S^{best} = \{1,2,3\}$ for $P > 34.85\,dB$. The user selection procedure of ZFS algorithm and $S^{best}$ at different transmit SNRs are listed in TABLE I.

According to TABLE I, the initially selected user $\{2\}$ is a redundant user for $S_{ZFS}$ when the transmit SNR is $27.13 < P \leq 34.85$. However, the transmit power of the user 2 is not zero. Taking $P = 27.14\,dB$ as an example, the transmit power distribution is $\lambda_1^{-1}p_1 = \lambda_3^{-1}p_3 = 22.42\,dB$ and $\lambda_2^{-1}p_2 = 22.26\,dB$, indicating that a redundant user exists even if $p_i > 0$ for each selected user. In fact, as we will show in Section VI-B, the case of redundant users with $p_i = 0$ does not exist when the ZFS algorithm is utilized to determine the user set.

2. Deleting users with $p_i = 0$ cannot guarantee the maximum sum rate increment

Which user should be deleted when redundant users exist in the selected user set? An intuitive method is to delete the user with the smallest effective-channel-gain $\lambda_i$, which corresponds to the user with $p_i = 0$ when a non-positive power allocation exists. However, the sum rate is affected by transmit SNR, channel norm, and channel correlation of selected users while the effective-channel-gain $\lambda_i$ only represents partial influence of channel norm and channel correlation. We have the following lemma.

**Lemma 1**: When a redundant user exists in the selected user set, deleting the user with $p_i = 0$ increases the sum rate but cannot guarantee the maximum sum rate increment.

*Proof:* See the Appendix. □

### B. Local optimum $S_n \neq U_n$

Define the neighborhood of $S_n$ as the set obtained by adding or deleting one user from $S_n$. The output of ZFS may fall into a local optimum, i.e., the sum rate of $S_{ZFS}$ cannot be increased by adding or deleting one user but is still not the global optimum. As shown in Fig. 2 and TABLE I, when $10.48\,dB < P \leq 27.13\,dB$ we have $S_{_{ZFS}} = \{1, 2\}$ and $S^{best} = \{1, 3\}$. The sum rate of $S_{_{ZFS}} = \{1, 2\}$ cannot increase by adding a new user 3 or by deleting the selected user 1 or 2, but $S_{_{ZFS}} \neq S^{best}$. We noticed, however, that the global optimum $S^{best}$ can be achieved from $S_{_{ZFS}}$ by swapping user 2 with user 3.

We can leave the local optimum through 'swap' operation on the user set $S_{_{ZFS}}$. However, there is a tradeoff between complexity and performance on the selection of 'swap' operation. When all possible 'swap's are allowed (one-for-one, one-for-many and many-for-one), the complexity is the same as exhaustive search. In this work, for the simplicity of implementation we considered only the one-for-one swap. Although it cannot guarantee the global optimum, the complexity will be greatly reduced. And we will show later that in most cases the sum rate optimum can be achieved by using one-for-one swapping.

According to the above analysis, to solve the flaws of traditional incremental greedy user selection algorithm we need 'delete' and 'swap' operations on the selected user set. Determining the best user to 'delete' or the best user pair to 'swap' requires sum rate comparison among all possible deleted or swapped user sets. According to (10)-(13), calculating the sum rate involves a Moore-Penrose pseudo-inverse which brings significant amount of complexity. In order to reduce the algorithm complexity, the recursive of $(\mathbf{H}_S \mathbf{H}_S^*)^{-1}$ was used in [2] and the LQ decomposition of $\mathbf{H}_S$ was used in [5] to calculate the effective-channel-gain $\lambda$ and the sum rate without calculating Moore-Penrose pseudo-inverse. However, the iteration methods in [2] and [5] only support adding a new user to the selected user set; they cannot be expanded to calculate the new sum rate when 'delete' or 'swap' operation is utilized. So, we need a new $\lambda$ updating method which can be used to calculate new sum rate after 'add', 'delete' and 'swap' operation while maintaining the same level of complexity. A new user selection algorithm will be constructed by using the new $\lambda$ updating method in Section V.

## IV. $\lambda$ Updating Method Based on ECV

According to (10)-(13), the effective-channel-gain $\lambda$ is the key parameter in calculating the sum rate of selected user set $S$. All the previous complexity reduction methods in ZFS and SWF update $\lambda$ through iteratively updating $\mathbf{H}_S^{\dagger}$ and are only applicable when a new user is added to $S$. To construct a method suitable for 'add', 'delete' and 'swap' operation, we designed an efficient $\lambda$ updating strategy that is based on iteratively updating ECV $\boldsymbol{\nu}$ defined in (8) instead of $\mathbf{H}_S^{\dagger}$ to reduce the complexity.

Let $U = \{1, \cdots, K\}$ be the index set of all users and $S$ be the index set of the selected user set. The proposed $\lambda$ updating strategy involves two classes of parameters, which correspond to the users in $S$ and $U \setminus S$ respectively, as follows:

- The ECV $\boldsymbol{\nu}_i$ of the selected user $i \in S$, according to (5) and (8) we have

$$\boldsymbol{\nu}_i = \mathbf{h}_i \left(\mathbf{I}_M - \mathbf{H}_{S\setminus\{i\}}^{*}(\mathbf{H}_{S\setminus\{i\}}\mathbf{H}_{S\setminus\{i\}}^{*})^{-1}\mathbf{H}_{S\setminus\{i\}}\right). \tag{18}$$

- The orthogonal component of channel vectors $\mathbf{g}_i$ of the remain user $j \in U \setminus S$, which is orthogonal to the subspace spanned by the channels of the selected users, where

$$\mathbf{g}_j = \mathbf{h}_j \left(\mathbf{I}_M - \mathbf{H}_S^{*}(\mathbf{H}_S\mathbf{H}_S^{*})^{-1}\mathbf{H}_S\right). \tag{19}$$

We need to update these two classes of parameters after each 'add', 'delete' and 'swap' operation to get the new effective-channel-gain $\lambda = \|\boldsymbol{\nu}\|^2$ for the new user set. The updating strategies of $\boldsymbol{\nu}_i$ and $\mathbf{g}_i$ under three operations are illustrated from both algebra and geometry perspectives in the following.

### *A. Add a new user*

Suppose a new user $k \in U \setminus S$ is added into the selected user set $S$, and denote the new user set as $S^+$ where $S^+ = S \cup \{k\}$. The ECV $\boldsymbol{\nu}_i$ of the selected users $i \in S$ and the $\mathbf{g}_i$ of the remaining users $j \in U \setminus S$ are known. We need to calculate the updated $\boldsymbol{\nu}_i^+$ of users $i \in S^+$ and $\mathbf{g}_i^+$ of users $j \in U \setminus S^+$.

1) Update $\boldsymbol{\nu}_i^+$

Since $S^+ \setminus \{k\} = (S \cup \{k\}) \setminus \{k\} = S$, the ECV of the new added user $k$ can be calculated according to (18)(19) as

$$\begin{aligned}\boldsymbol{\nu}_k^+ &= \mathbf{h}_k \left(\mathbf{I}_M - \mathbf{H}_{S^+\setminus\{k\}}^{*}(\mathbf{H}_{S^+\setminus\{k\}}\mathbf{H}_{S^+\setminus\{k\}}^{*})^{-1}\mathbf{H}_{S^+\setminus\{k\}}\right) \\ &= \mathbf{h}_k \left(\mathbf{I}_M - \mathbf{H}_S^{*}(\mathbf{H}_S\mathbf{H}_S^{*})^{-1}\mathbf{H}_S\right) \\ &= \mathbf{g}_k \,.\end{aligned} \tag{20}$$

As for the other users $i \in S^+ \setminus \{k\}$, or $i \in S$, we have

$$\mathbf{H}_{S^+\setminus\{i\}} = \mathbf{H}_{S\setminus\{i\}\cup\{k\}} = \begin{bmatrix} \mathbf{H}_{S\setminus\{i\}} \\ \mathbf{h}_k \end{bmatrix}.$$

After some algebraic manipulation, we obtain

$$\begin{aligned}
\boldsymbol{\nu}_i^+ &= \mathbf{h}_i \left(\mathbf{I}_M - \mathbf{H}^*_{S^+\setminus\{i\}}(\mathbf{H}_{S^+\setminus\{i\}}\mathbf{H}^*_{S^+\setminus\{i\}})^{-1}\mathbf{H}_{S^+\setminus\{i\}}\right) \\
&= \mathbf{h}_i \left(\mathbf{I}_M - \left[\begin{array}{cc} \mathbf{H}^*_{S\setminus\{i\}} & \mathbf{h}_k^* \end{array}\right] \left[\begin{array}{cc} \mathbf{H}_{S\setminus\{i\}}\mathbf{H}^*_{S\setminus\{i\}} & \mathbf{H}_{S\setminus\{i\}}\mathbf{h}_k^* \\ \mathbf{h}_k\mathbf{H}^*_{S\setminus\{i\}} & \mathbf{h}_k\mathbf{h}_k^* \end{array}\right]^{-1} \left[\begin{array}{c} \mathbf{H}_{S\setminus\{i\}} \\ \mathbf{h}_k \end{array}\right]\right) \\
&= \boldsymbol{\nu}_i \left(\mathbf{I} - \frac{\mathbf{h}_k^*\boldsymbol{\nu}_k(S^+\setminus\{i\})}{\|\boldsymbol{\nu}_k(S^+\setminus\{i\})\|^2}\right) ,
\end{aligned} \tag{21}$$

where $\boldsymbol{\nu}_k(S^+\setminus\{i\}) = \mathbf{h}_k \left(\mathbf{I}_M - \mathbf{H}^*_{S\setminus\{i\}}(\mathbf{H}_{S\setminus\{i\}}\mathbf{H}^*_{S\setminus\{i\}})^{-1}\mathbf{H}_{S\setminus\{i\}}\right)$ is the ECV of user $k$ when the selected user set is $S^+\setminus\{i\}$. Since

$$\begin{aligned}
\mathbf{g}_k &= \mathbf{h}_k \left(\mathbf{I}_M - \mathbf{H}_S^*(\mathbf{H}_S\mathbf{H}_S^*)^{-1}\mathbf{H}_S\right) \\
&= \mathbf{h}_k \left(\mathbf{I}_M - \left[\begin{array}{cc} \mathbf{H}^*_{S\setminus\{i\}} & \mathbf{h}_i^* \end{array}\right] \left[\begin{array}{cc} \mathbf{H}_{S\setminus\{i\}}\mathbf{H}^*_{S\setminus\{i\}} & \mathbf{H}_{S\setminus\{i\}}\mathbf{h}_i^* \\ \mathbf{h}_i\mathbf{H}^*_{S\setminus\{i\}} & \mathbf{h}_i\mathbf{h}_i^* \end{array}\right]^{-1} \left[\begin{array}{c} \mathbf{H}_{S\setminus\{i\}} \\ \mathbf{h}_i \end{array}\right]\right) \\
&= \boldsymbol{\nu}_k(S^+\setminus\{i\}) - \frac{\mathbf{h}_k\boldsymbol{\nu}_i^*}{\|\boldsymbol{\nu}_i\|^2}\nu_i ,
\end{aligned} \tag{22}$$

according to (22) we have

$$\boldsymbol{\nu}_k(S^+\setminus\{i\}) = \mathbf{g}_k + \frac{\mathbf{h}_k\boldsymbol{\nu}_i^*}{\|\boldsymbol{\nu}_i\|^2}\boldsymbol{\nu}_i . \tag{23}$$

Plugging (23) into (21), we get

$$\begin{aligned}
\boldsymbol{\nu}_i^+ &= \boldsymbol{\nu}_i \left(\mathbf{I} - \frac{\mathbf{h}_k^*\left(\mathbf{g}_k + \frac{\mathbf{h}_k\boldsymbol{\nu}_i^*}{\|\boldsymbol{\nu}_i\|^2}\boldsymbol{\nu}_i\right)}{\left(\mathbf{g}_k + \frac{\mathbf{h}_k\boldsymbol{\nu}_i^*}{\|\boldsymbol{\nu}_i\|^2}\boldsymbol{\nu}_i\right)\mathbf{h}_k^*}\right) \\
&= \frac{\|\boldsymbol{\nu}_i\|^2\|\mathbf{g}_k\|^2}{\|\boldsymbol{\nu}_i\|^2\|\mathbf{g}_k\|^2 + \|\boldsymbol{\nu}_i\mathbf{h}_k^*\|^2}\left(\boldsymbol{\nu}_i - \frac{\boldsymbol{\nu}_i\mathbf{h}_k^*}{\|\mathbf{g}_k\|^2}\mathbf{g}_k\right) .
\end{aligned} \tag{24}$$

Since $\mathbf{g}_k \perp \boldsymbol{\nu}_i$, the effective-channel-gain $\lambda_i^+$ is

$$\lambda_i^+ = \left\|\boldsymbol{\nu}_i^+\right\|^2 = \frac{\|\boldsymbol{\nu}_i\|^4\|\mathbf{g}_k\|^2}{\|\boldsymbol{\nu}_i\|^2\|\mathbf{g}_k\|^2 + \|\boldsymbol{\nu}_i\mathbf{h}_k^*\|^2} . \tag{25}$$

As shown in Fig. 3, the derivation of $\boldsymbol{\nu}_i^+$ from (21) to (25) can also be explained from geometry perspective. Since $\boldsymbol{\nu}_i^+$ is the component of $\mathbf{h}_i$ orthogonal to the subspace $V_i^+ = span\{\mathbf{h}_j|j \in S^+, j \neq i\}$, and $\boldsymbol{\nu}_i$ and $\boldsymbol{\nu}_k^+$ are orthogonal to the subspace $V_i = span\{\mathbf{h}_j|j \in S, j \neq i\}$, $\boldsymbol{\nu}_i^+$ can be calculated by the component of $\boldsymbol{\nu}_i$ orthogonal to $\boldsymbol{\nu}_k(S^+\setminus\{i\})$, which is the projection of $\mathbf{h}_k$ on the subspace $span\{\boldsymbol{\nu}_i, \boldsymbol{\nu}_k^+\}$ as shown in Fig. 3. Note that $span\{\boldsymbol{\nu}_i, \boldsymbol{\nu}_k^+\}$ is the subspace orthogonal to $V_i = span\{\mathbf{h}_j|j \in S, j \neq i\}$; $\boldsymbol{\nu}_i$ and $\boldsymbol{\nu}_k(S^+\setminus\{i\})$ are the orthogonal components of $\mathbf{h}_i$ and $\mathbf{h}_k$ projected onto the subspace $V_i$. Supposing the angle between $\boldsymbol{\nu}_i$ and $\boldsymbol{\nu}_i^+$ is $\theta$, we have

$$\cos\theta = \frac{\|\boldsymbol{\nu}_k^+\|}{\|\boldsymbol{\nu}_k(S^+ \setminus \{i\})\|} = \sqrt{\frac{\|\boldsymbol{\nu}_i\|^2 \|\mathbf{g}_k\|^2}{\|\boldsymbol{\nu}_i\|^2 \|\mathbf{g}_k\|^2 + \|\boldsymbol{\nu}_i \mathbf{h}_k^*\|^2}} \tag{26}$$

$$\lambda_i^+ = \|\boldsymbol{\nu}_i\|^2 \cos^2\theta = \lambda_i \cos^2\theta \tag{27}$$

$$\boldsymbol{\nu}_i^+ = \left(\boldsymbol{\nu}_i - \frac{\boldsymbol{\nu}_i \mathbf{h}_k^*}{\|\mathbf{g}_k\|^2}\mathbf{g}_k\right)\cos^2\theta\,. \tag{28}$$

2) Update $\mathbf{g}_j^+$

According to (19), we can calculate the updated $\mathbf{g}_j^+$ with the same method as in (21)-(24) for the users $j \in U \setminus S^+$. However, since $\mathbf{g}_j^+$ is the component of $\mathbf{h}_j$ orthogonal to the subspace $V^+ = span\{\mathbf{h}_i | i \in S^+\}$ and $\mathbf{g}_j$ is orthogonal to the subspace $V = span\{\mathbf{h}_i | i \in S\}$, we can find $\mathbf{g}_j^+$ via Gram-Schmidt orthogonal procedure by projecting $\mathbf{g}_j$ onto orthogonal complement of the vector $\mathbf{u}$, where $\mathbf{u} \perp V$ and $V^+ = span\{V, \mathbf{u}\}$. According to former analysis, $\mathbf{u} = \boldsymbol{\nu}_k^+ = \mathbf{g}_k$, so

$$\mathbf{g}_j^+ = \mathbf{g}_j - \frac{\mathbf{g}_j \mathbf{g}_k^*}{\|\mathbf{g}_k\|^2}\mathbf{g}_k\,. \tag{29}$$

for the users $j \in U \setminus S^+$.

In summary, the updated $\boldsymbol{\nu}_i^+$ of users $i \in S^+$ and $\mathbf{g}_j^+$ of users $j \in U \setminus S^+$ are listed as follows:

$$\boldsymbol{\nu}_i^+ = \begin{cases} \frac{\lambda_i \|\mathbf{g}_k\|^2}{\lambda_i \|\mathbf{g}_k\|^2 + \|\boldsymbol{\nu}_i \mathbf{h}_k^*\|^2}\left(\boldsymbol{\nu}_i - \frac{\boldsymbol{\nu}_i \mathbf{h}_k^*}{\|\mathbf{g}_k\|^2}\mathbf{g}_k\right), & i \in S \\ \mathbf{g}_k, & i = k \end{cases} \tag{30}$$

$$\mathbf{g}_j^+ = \mathbf{g}_j - \frac{\mathbf{g}_j \mathbf{g}_k^*}{\|\mathbf{g}_k\|^2}\mathbf{g}_k, \quad j \in U \setminus S \setminus \{k\}\,. \tag{31}$$

### *B. Delete a selected user*

Suppose the user $k \in S$ is deleted from the selected user set $S$, and denote the new user set as $S^-$ where $S^- = S \setminus \{k\}$. We need to calculate the updated $\boldsymbol{\nu}_i^-$ for users $i \in S^-$ and updated $\mathbf{g}_i^-$ for users $j \in U \setminus S^-$.

1) Update $\boldsymbol{\nu}_i^-$

The ECV $\boldsymbol{\nu}_i^-$ is the component of $\mathbf{h}_i$ that is orthogonal to the subspace $V_i^- = span\{\mathbf{h}_j | j \in S^-, j \neq i\}$. Since $\boldsymbol{\nu}_i \perp V_i$ and $\boldsymbol{\nu}_k \perp V_i^-$ , where $V_i = span\{\mathbf{h}_j | j \in S, j \neq i\} = span\{V_i^-, \boldsymbol{\nu}_k\}$, the ECV $\boldsymbol{\nu}_i^-$ can be expressed as the projection of $\mathbf{h}_i$ on the subspace $span\{\boldsymbol{\nu}_i, \boldsymbol{\nu}_k\}$. This is equivalent to solving $\boldsymbol{\nu}_i$ when knowing $\boldsymbol{\nu}_i^+$ and $\boldsymbol{\nu}_k^+$ in Fig. 3, where $\boldsymbol{\nu}_i$ is the projection of $\mathbf{h}_i$ on the subspace $span\{\boldsymbol{\nu}_i^+, \boldsymbol{\nu}_k^+\}$. Thus, we have [18]

$$\begin{aligned}
\boldsymbol{\nu}_i^- &= \mathbf{h}_i \begin{bmatrix} \boldsymbol{\nu}_i^* & \boldsymbol{\nu}_k^* \end{bmatrix} \begin{bmatrix} \boldsymbol{\nu}_i\boldsymbol{\nu}_i^* & \boldsymbol{\nu}_i\boldsymbol{\nu}_k^* \\ \boldsymbol{\nu}_k\boldsymbol{\nu}_i^* & \boldsymbol{\nu}_k\boldsymbol{\nu}_k^* \end{bmatrix}^{-1} \begin{bmatrix} \boldsymbol{\nu}_i \\ \boldsymbol{\nu}_k \end{bmatrix} \\
&= \begin{bmatrix} \mathbf{h}_i\boldsymbol{\nu}_i^* & 0 \end{bmatrix} \frac{1}{\|\boldsymbol{\nu}_i\|^2 \|\boldsymbol{\nu}_k\|^2 - \|\boldsymbol{\nu}_i\boldsymbol{\nu}_k^*\|^2} \begin{bmatrix} \boldsymbol{\nu}_k\boldsymbol{\nu}_k^* & -\boldsymbol{\nu}_i\boldsymbol{\nu}_k^* \\ -\boldsymbol{\nu}_k\boldsymbol{\nu}_i^* & \boldsymbol{\nu}_i\boldsymbol{\nu}_i^* \end{bmatrix} \begin{bmatrix} \boldsymbol{\nu}_i \\ \boldsymbol{\nu}_k \end{bmatrix} \\
&= \frac{\|\boldsymbol{\nu}_i\|^2 \|\boldsymbol{\nu}_k\|^2}{\|\boldsymbol{\nu}_i\|^2 \|\boldsymbol{\nu}_k\|^2 - \|\boldsymbol{\nu}_i\boldsymbol{\nu}_k^*\|^2} \left( \boldsymbol{\nu}_i - \frac{\boldsymbol{\nu}_i\boldsymbol{\nu}_k^*}{\|\boldsymbol{\nu}_k\|^2}\boldsymbol{\nu}_k \right) .
\end{aligned} \tag{32}$$

The second equality holds because $\boldsymbol{\nu}_k \perp V_k$, where $V_k = span\{\mathbf{h}_j | j \in S, j \neq k\}$, thus, $\mathbf{h}_i\boldsymbol{\nu}_k^* = 0$. The third equality holds because $\mathbf{h}_i\boldsymbol{\nu}_i^* = \mathbf{h}_i(\mathbf{P}_i^\perp)^*\mathbf{h}_i^* = \mathbf{h}_i\mathbf{P}_i^\perp(\mathbf{P}_i^\perp)^*\mathbf{h}_i^* = \|\boldsymbol{\nu}_i\|^*$, where $\mathbf{P}_i^\perp = \mathbf{I}_M - \mathbf{H}_{S\setminus\{i\}}^*(\mathbf{H}_{S\setminus\{i\}}\mathbf{H}_{S\setminus\{i\}}^*)^{-1}\mathbf{H}_{S\setminus\{i\}}$ is an idempotent Hermitian matrix that $\left(\mathbf{P}_i^\perp\right)^2 = \mathbf{P}_i^\perp$ and $\left(\mathbf{P}_i^\perp\right)^* = \mathbf{P}_i^\perp$.

According to (32), the effective-channel-gain $\lambda_i^-$ for users $i$ is

$$\lambda_i^- = \left\|\boldsymbol{\nu}_i^-\right\|^2 = \lambda_i \frac{\|\boldsymbol{\nu}_i\|^2 \|\boldsymbol{\nu}_k\|^2}{\|\boldsymbol{\nu}_i\|^2 \|\boldsymbol{\nu}_k\|^2 - \|\boldsymbol{\nu}_i\boldsymbol{\nu}_k^*\|^2} . \tag{33}$$

The above deduction for $\boldsymbol{\nu}_i^-$ can also be explained from the geometry perspective as shown in Fig. 4. The $\boldsymbol{\nu}_i^-$ is in the subspace $span\{\boldsymbol{\nu}_i, \boldsymbol{\nu}_k\}$ and orthogonal to $\boldsymbol{\nu}_k$. Suppose the angle between $\boldsymbol{\nu}_i^-$ and $\boldsymbol{\nu}_i$ is $\theta$, we have

$$\cos\theta = \sqrt{1-\sin^2\theta} = \sqrt{1 - \frac{\|\boldsymbol{\nu}_i\boldsymbol{\nu}_k^*\|^2}{\|\boldsymbol{\nu}_i\|^2 \|\boldsymbol{\nu}_k\|^2}} \tag{34}$$

$$\lambda_i^- = \|\boldsymbol{\nu}_i\|^2 \cos^{-2}\theta = \lambda_i \cos^{-2}\theta \tag{35}$$

$$\boldsymbol{\nu}_i^- = \left( \boldsymbol{\nu}_i - \frac{\boldsymbol{\nu}_i\boldsymbol{\nu}_k^*}{\|\boldsymbol{\nu}_k\|^2}\boldsymbol{\nu}_k \right) \cos^{-2}\theta . \tag{36}$$

2) Update $\mathbf{g}_i^-$

The deleted user $k$ is now moved from the previously selected user set $S$ to the remaining user set $U \setminus S^-$. Since $S^- = S \setminus \{k\}$, $\mathbf{g}_k^-$ can be calculated according to (18)(19) as

$$\begin{aligned}
\mathbf{g}_k^- &= \mathbf{h}_k \left( \mathbf{I}_M - \mathbf{H}_{S^-}^*(\mathbf{H}_{S^-}\mathbf{H}_{S^-}^*)^{-1}\mathbf{H}_{S^-} \right) \\
&= \mathbf{h}_k \left( \mathbf{I}_M - \mathbf{H}_{S\setminus\{k\}}^*(\mathbf{H}_{S\setminus\{k\}}\mathbf{H}_{S\setminus\{k\}}^*)^{-1}\mathbf{H}_{S\setminus\{k\}} \right) \\
&= \boldsymbol{\nu}_k .
\end{aligned} \tag{37}$$

As for the other users $j \in (U \setminus S^-) \setminus \{k\}$, or $j \in U \setminus S$, we can update the $\mathbf{g}_j^-$ according to Gram-Schmidt orthogonal procedure. Since $\mathbf{g}_j^-$ is the component of $\mathbf{h}_j$ orthogonal to the subspace $V^- = span\{\mathbf{h}_i | i \in S^-\}$, $\mathbf{g}_i \perp V$ and $\boldsymbol{\nu}_k \perp V^-$, where $V = span\{\mathbf{h}_j | j \in S\} = span\{V^-, \boldsymbol{\nu}_k\}$, the updated $\mathbf{g}_j^-$ can be expressed as the combination of $\mathbf{g}_i$ and the projection of $\mathbf{h}_i$ on $\boldsymbol{\nu}_k$, i.e.,

$$\mathbf{g}_j^- = \mathbf{g}_j + \frac{\mathbf{h}_j \boldsymbol{\nu}_k^*}{\|\boldsymbol{\nu}_k\|^2} \boldsymbol{\nu}_k \tag{38}$$

for the users $j \in U \setminus S$.

In summary, the updated $\boldsymbol{\nu}_i^-$ of users $i \in S^-$ and $\mathbf{g}_j^-$ of users $j \in U \setminus S^-$ are listed as following:

$$\boldsymbol{\nu}_i^- = \frac{\lambda_i \lambda_k}{\lambda_i \lambda_k - \|\boldsymbol{\nu}_i \boldsymbol{\nu}_k^*\|^2} \left( \boldsymbol{\nu}_i - \frac{\boldsymbol{\nu}_i \boldsymbol{\nu}_k^*}{\lambda_k} \boldsymbol{\nu}_k \right), \quad i \in S \setminus \{k\} \tag{39}$$

$$\mathbf{g}_j^- = \begin{cases} \mathbf{g}_j + \frac{\mathbf{h}_j \boldsymbol{\nu}_k^*}{\lambda_k} \boldsymbol{\nu}_k, & j \in U \setminus S \\ \boldsymbol{\nu}_k, & j = k \end{cases}. \tag{40}$$

*C. Swap users one-for-one*

Suppose a new user $l \in U \setminus S$ is swapped with a selected user $k \in S$, and denote the new user set as $S^S$ where $S^S = (S \cup \{l\}) \setminus \{k\}$. We need to calculate the updated $\boldsymbol{\nu}_i^s$ for users $i \in S^S$ and updated $\mathbf{g}_j^s$ for users $j \in U \setminus S^S$.

Since the one-for-one user swap is a combination of adding a new user and deleting a selected user, the corresponding $\boldsymbol{\nu}_i^s$ and $\mathbf{g}_j^s$ updating algorithm can be obtained by sequentially applying the ‘add’ and ‘delete’ updating algorithm, as defined in (30)(31) and (39)(40). Assume adding user $l$ first and then deleting user $k$. Denoting the intermediate results as $\boldsymbol{\nu}_{i+}$ and $\mathbf{g}_{j+}$, we have

$$\boldsymbol{\nu}_i^s = \frac{\|\boldsymbol{\nu}_{i+}\|^2 \|\boldsymbol{\nu}_{k+}\|^2}{\|\boldsymbol{\nu}_{i+}\|^2 \|\boldsymbol{\nu}_{k+}\|^2 - \left\|\boldsymbol{\nu}_{i+} \boldsymbol{\nu}_{k+}^*\right\|^2} \left( \boldsymbol{\nu}_{i+} - \frac{\boldsymbol{\nu}_{i+} \boldsymbol{\nu}_{k+}^*}{\|\boldsymbol{\nu}_{k+}\|^2} \boldsymbol{\nu}_{k+} \right), \quad i \in S^S \tag{41}$$

$$\lambda_i^s = \frac{\|\boldsymbol{\nu}_{i+}\|^4 \|\boldsymbol{\nu}_{k+}\|^2}{\|\boldsymbol{\nu}_{i+}\|^2 \|\boldsymbol{\nu}_{k+}\|^2 - \left\|\boldsymbol{\nu}_{i+} \boldsymbol{\nu}_{k+}^*\right\|^2}, \quad i \in S^S \tag{42}$$

$$\mathbf{g}_j^s = \begin{cases} \mathbf{g}_{j+} + \frac{\mathbf{h}_{j+} \boldsymbol{\nu}_{k+}^*}{\|\boldsymbol{\nu}_{k+}\|^2} \boldsymbol{\nu}_{k+}, & j \in (U \setminus S^S) \setminus \{k\} \\ \boldsymbol{\nu}_{k+}, & j = k \end{cases}. \tag{43}$$

where

$$\boldsymbol{\nu}_{i+} = \begin{cases} \frac{\lambda_i \|\mathbf{g}_l\|^2}{\lambda_i \|\mathbf{g}_l\|^2 + \left\|\boldsymbol{\nu}_i \mathbf{h}_l^*\right\|^2} \left(\boldsymbol{\nu}_i - \frac{\boldsymbol{\nu}_i \mathbf{h}_l^*}{\|\mathbf{g}_l\|^2} \mathbf{g}_l\right), & i \in S \\ \mathbf{g}_l \,, & i = l \end{cases} \tag{44}$$

$$\mathbf{g}_{j+} = \mathbf{g}_j - \frac{\mathbf{g}_j \mathbf{g}_l^*}{\|\mathbf{g}_l\|^2} \mathbf{g}_l, \quad j \in (U \setminus S) \setminus \{l\} \,. \tag{45}$$

Note: we can also get the same $\boldsymbol{\nu}_i^s$ and $\mathbf{g}_j^s$ by first deleting user $k$ and then adding user $l$. The expressions are similar to (41)-(45) with the same complexity and thus omitted for the sake of space.

## V. GUSS Algorithm

A new greedy user selection algorithm, which utilizes the ECV-based $\lambda$ updating strategy in Section IV, is proposed in this section. The algorithm is called greedy user selection with swap (GUSS) algorithm as it includes ‘add’, ‘delete’ and ‘swap’ operations.

The GUSS algorithm works as follows: it initializes with ZFS, i.e., adding one user with the maximal $\Delta R$ in each step consecutively until the maximal $\Delta R \leq 0$; it then deletes one user at a time, each deletion produces maximal $\Delta R$, until no sum rate increment is possible. GUSS oscillates between ‘sequential add’ and ‘sequential delete’ until $\Delta R \leq 0$ for both operation. One ‘swap’ operation is then invoked to boost the sum rate. After the ‘swap’, GUSS goes back to the oscillation of ‘add’ and ‘delete’, attempting to further increase the sum rate. If $\Delta R \leq 0$ for any user choice, the user selection procedure finishes. The construction and complexity analysis of GUSS algorithm are outlined next.

### *A. Construction of GUSS algorithm*

Let $U = \{1, \cdots, K\}$ be the index set of all users and $S$ be the index set of the selected user set. The $\boldsymbol{\nu}_i$ and $\lambda_i$ are the ECV and effective-channel-gain of selected user $i \in S$, and $\mathbf{g}_j$ for $j \in U \setminus S$ is the component of remaining channel vectors orthogonal to the subspace $span\{\mathbf{h}_i | i \in S\}$.

**Step 1) Initialization:**

$$S = \emptyset$$

$$\mathbf{g}_j = \mathbf{h}_j \text{ for all user } j \in U \,.$$

**Step 2) Add a new user:**

$$\lambda_i^+(w) = \begin{cases} \frac{\lambda_i^2 \|\mathbf{g}_w\|^2}{\lambda_i \|\mathbf{g}_w\|^2 + \|\boldsymbol{\nu}_i \mathbf{h}_w^*\|^2}, & i \in S \\ \|\mathbf{g}_w\|^2, & i = w \end{cases} \tag{46}$$

$$k = \arg \max_{w \in U \setminus S} R(S \cup \{w\}) \,. \tag{47}$$

Let $\Delta R = R(S \cup \{k\}) - R(S)$. If $\Delta R > 0$, $S \leftarrow S \cup \{k\}$, update $\boldsymbol{\nu}_i$, $\mathbf{g}_j$ and corresponding $\lambda_i$ according to (30)(31) and then go to step 2); if $\Delta R \leq 0$ for one iteration, go to step 3); else if $\Delta R \leq 0$ for two consecutive iterations, go to step 4).

**Step 3) Delete a selected user:**

$$\lambda_i^-(w) = \frac{\lambda_i^2 \lambda_w}{\lambda_i \lambda_w - \|\boldsymbol{\nu}_i \boldsymbol{\nu}_w^*\|^2}, \quad i \in S \setminus \{w\} \tag{48}$$

$$k = \arg\max_{w \in S} R(S \setminus \{w\}) \,. \tag{49}$$

Let $\Delta R = R(S \setminus \{k\}) - R(S)$. If $\Delta R > 0$, $S \leftarrow S \setminus \{k\}$, update $\boldsymbol{\nu}_i$, $\mathbf{g}_j$ and corresponding $\lambda_i$ according to (39)(40) and then go to step 3); if $\Delta R \leq 0$ for one iteration, go to step 2); else if $\Delta R \leq 0$ for two consecutive iterations, go to step 4).

**Step 4) Swap users one-for-one:**

$$\lambda_i^s(k,l) = \frac{\left\|\boldsymbol{\nu}_{i+,l}\right\|^4 \left\|\boldsymbol{\nu}_{k+,l}\right\|^2}{\left\|\boldsymbol{\nu}_{i+,l}\right\|^2 \left\|\boldsymbol{\nu}_{k+,l}\right\|^2 - \left\|\boldsymbol{\nu}_{i+,l} \boldsymbol{\nu}_{k+,l}^*\right\|^2}, \quad i \in S \cup \{l\} \setminus \{k\} \tag{50}$$

$$\boldsymbol{\nu}_{i+,l} = \begin{cases} \frac{\lambda_i \|\mathbf{g}_l\|^2}{\lambda_i \|\mathbf{g}_l\|^2 + \left\|\boldsymbol{\nu}_i \mathbf{h}_l^*\right\|^2} \left( \boldsymbol{\nu}_i - \frac{\boldsymbol{\nu}_i \mathbf{h}_l^*}{\left\|\mathbf{g}_l\right\|^2} \mathbf{g}_l \right), & i \in S \\ \mathbf{g}_l, & i = l \end{cases} \tag{51}$$

$$\{k, l\} = \arg \max_{k \in S, \, l \in U \setminus S} R(S \cup \{l\} \setminus \{k\}) \,. \tag{52}$$

Let $\Delta R = R(S \cup \{l\} \setminus \{k\}) - R(S)$. If $\Delta R > 0$, $S \leftarrow S \cup \{l\} \setminus \{k\}$, update $\boldsymbol{\nu}_i$, $\mathbf{g}_j$ and corresponding $\lambda_i$ according to (41)-(43) and then go to step 2); if $\Delta R \leq 0$, go to step 5).

**Step 5) Precoding matrix:**

$$\left[ \ \frac{\sqrt{\mu \lambda_{(1)} - 1}}{\lambda_{(1)}} \boldsymbol{\nu}_{(1)}^*, \quad \frac{\sqrt{\mu \lambda_{(2)} - 1}}{\lambda_{(2)}} \boldsymbol{\nu}_{(2)}^*, \quad \cdots, \quad \frac{\sqrt{\mu \lambda_{(n)} - 1}}{\lambda_{(n)}} \boldsymbol{\nu}_{(n)}^* \ \right], \tag{53}$$

where $n = |S|$, $\boldsymbol{\nu}_{(i)}$ and $\lambda_{(i)}$ are the ECV and effective-channel-gain of the $i$-th user in $S$, and $\mu = \left(P + \sum_{i \in S} \lambda_i^{-1}\right)/n$ is the water level for power allocation.

GUSS initializes with empty user set $S = \emptyset$. The first selected user is the one with the maximal effective-channel-gains $\lambda_i^+(w)$ which is equivalent to the maximal square channel norm $\|h_i\|^2$ for $S = \emptyset$. GUSS repeats the add operation in step 2) sequentially, and the procedure before it goes to step 3) for the first time constitutes the user selection of ZFS algorithm.

For each 'add', 'delete' and 'swap' operations in step2) to step 4), the updated effective-channel-gains is calculated first and then used to evaluate the updated sum rate with waterfilling power allocation. To further reduce the complexity, we can eliminate the iterative waterfilling procedure that is involved in (47)(49)(52) by restricting the candidate user or user pair to the ones that provide positive transmit power for all users in the updated user set. Take (47) in step 2) as an example, from the properties of waterfilling, this holds if [2]

$$\frac{|S|+1}{\min_{i\in S\cup\{w\}} \lambda_i^+(w)} < P + \sum_{i\in S\cup\{w\}} \frac{1}{\lambda_i^+(w)}\,. \tag{54}$$

If (54) is satisfied, the corresponding water level can be calculated directly through

$$\mu = \frac{1}{|S|+1}\left(P + \sum_{i\in S\cup\{w\}} \frac{1}{\lambda_i^+(w)}\right)\,. \tag{55}$$

Similar inequality can be achieved for step 3) and 4). According to our simulations, this search space pruning operation does not compromise sum rate at all.

The searching space in step 4) can be further reduced by $K-|S|$ or $|S|$, because it provides non-positive sum rate increment if the last added or deleted user is involved in the one-for-one swap. The calculation of all $\lambda_i^s(k,l)$ in (50) involves $K-|S|$ partial $\boldsymbol{\nu}_{i+,l}$s updates as in (51) if it adds user $l$ first. If we calculate the $\lambda_i^s(k,l)$ by first deleting user $k$, the corresponding $\lambda_i^s(k,l)$ updating involves $|S|$ partial $\boldsymbol{\nu}_{i-,k}$s and $\mathbf{g}_{j-,k}$s updates as

$$\lambda_i^s(k,l) = \begin{cases} \dfrac{\|\boldsymbol{\nu}_{i-,k}\|^4\|\mathbf{g}_{l-,k}\|^2}{\|\boldsymbol{\nu}_{i-,k}\|^2\|\mathbf{g}_{l-,k}\|^2+\|\boldsymbol{\nu}_{i-,k}\mathbf{h}_l^*\|^2}, & i\in S\setminus\{k\} \\ \|\mathbf{g}_{l-,k}\|^2, & i=l \end{cases} \tag{56}$$

$$\boldsymbol{\nu}_{i-,k} = \frac{\lambda_i\lambda_k}{\lambda_i\lambda_k-\lambda_k}\left(\boldsymbol{\nu}_i - \frac{\boldsymbol{\nu}_i\boldsymbol{\nu}_k^*}{\lambda_k}\boldsymbol{\nu}_k\right), \quad i\in S\setminus\{k\} \tag{57}$$

$$\mathbf{g}_{j-,k} = \mathbf{g}_j + \frac{\mathbf{h}_j\boldsymbol{\nu}_k^*}{\lambda_k}\boldsymbol{\nu}_k, \quad j\in U\setminus S\,. \tag{58}$$

In step 5), the precoding matrix in (53) is the result of ZFBF and waterfilling power allocation. According to (2), the precoding matrix can be written in the form

$$[\ w_{(1)}\sqrt{p_{(1)}},\ \cdots,\ w_{(n)}\sqrt{p_{(n)}}\ ]\,. \tag{59}$$

The transmit power scaling factor is

$$p_{(i)} = \mu\lambda_{(i)} - 1 > 0\,. \tag{60}$$

for all users because all the selected users of GUSS will be allocated with positive transmit power. If not, the sum rate can be increased by 'delete' operation, which is contradictory to the fact that the user set output $S_{GUSS}$ of GUSS cannot be increased by 'add', 'delete' or 'one-for-one swap' operation. By plugging (9) and (60) into (59), we got the precoding matrix (53).

By construction, GUSS provides a sum rate higher than or equal to the one achieved by ZFS because the selected user set $S$ is improved by allowing 'delete' and 'swap' operations on the basis of ZFS. To distinguish the source of performance improvement, we constructed here another user selection algorithm that only allows 'add' and 'delete' operations, named greedy user selection without swap (GUS-nS) algorithm. GUS-nS removes the swap operation in step 4) of GUSS; therefore, the user selection process finishes if $\Delta R \leq 0$ for two consecutive iterations in step 2) or step 3). So, GUS-nS improves ZFS by only by eliminating the redundant users without handling the local optimum flaws.

### *B. Complexity analysis*

The computational complexity of the proposed algorithm includes two parts: 1) user search; and 2) $\boldsymbol{\nu}_i$, $\mathrm{g}_j$ and $\lambda_i$ update. We focused on the complexity of user search as $\boldsymbol{\nu}_i$, $\mathrm{g}_j$ and $\lambda_i$ updating stage has fixed complexity and is negligible when compared with user search. Let $n = |S|$ denote the cardinality of $S$. The complexity of each step is calculated as follows.

- For a given $S$ in step 2), the GUSS algorithm evaluates $K - n$ rates $R(S \cup \{w\})$. The evaluation of $R(S \cup \{w\})$ is split into the evaluation of $\lambda_i^+(w)$ followed by evaluation of $\mu$ according to (10). The evaluation of all $\lambda_i^+(w)$ for $i \in S \cup \{w\}$ requires $n$ vector-vector multiplications and $n+1$ vector 2-norms (vectors are $1 \times M$), and thus has $M(2n+1)$ multiplications. Repeating this over $K-n$ remain users, we obtain the user search complexity in step 2) as $M(K-n)(2n+1)$ multiplications.
- For a given $S$ in step 3), the GUSS algorithm evaluates $n$ rates $R(S \setminus \{w\})$. Similar to step 2), the evaluation of $\lambda_i^-(w)$ for $i \in S \setminus \{w\}$ involves $M(2n-1)$ multiplications. Repeating this over $n$ selected users, we obtain the user search complexity in step 3) as $Mn(2n-1)$ multiplications.

- For a given $S$ in step 4), the GUSS algorithm evaluates $Kn - n^2$ rates $R(S \cup \{l\} \setminus \{k\})$. Suppose $\lambda_i^s(k,l)$s are calculated according to (50)(51), i.e., 'add' precedes 'delete' in a 'swap'. The user search involves $2Mn^2+3Mn+M$ multiplications for each group $\lambda_i^s(k,l)$s with $k \in S$, and $(2Mn^2+3Mn+M)(K-n)$ complex multiplications for all. The user search involves $2MKn^2 - 2Mn^3 + 3Mn^2 - 3Mn$ complex multiplications if $\lambda_i^s(k,l)$s are calculated according to (56)-(58), i.e., 'delete' precedes 'add' in a 'swap'. However, they all have the same level of complexity $O\left(2Mn^2(K-n)\right)$.

The total complexity of GUSS in step 2) is approximately $\sum_{n=1}^{M} M(K-n)(2n+1)$, which is $O(KM^3 - \frac{2}{3}M^4)$. Suppose the number of iterations in step 3) and 4) is $b$ and $a$ respectively, which will be shown to be small numbers in next section. The total complexity of GUSS in step 3) and 4) are $O(2bM^3)$ and $O\left(2aKM^3 - 2aM^4\right)$. So, the complexity of GUSS is $O\left((2a+1)KM^3 - (2a+\frac{2}{3})M^4\right)$, and the complexity of GUS-nS is $O(KM^3 - \frac{2}{3}M^4)$. When the number of users $K \gg M$, the complexity of GUSS and GUS-nS is simplified as $O\left((2a+1)KM^3\right)$ and $O\left(KM^3\right)$, respectively. Since the complexity of both ZFS and SWF is $O\left(KM^3\right)$, the GUS-nS has the same complexity with ZFS and SWF, and GUSS has $2a+1$ linear complexity increment. However, as it will be shown in next section, both GUSS and GUS-nS outperform ZFS and SWF in terms of achieved sum rate.

## VI. Simulation Resultes

In this section, we present the numerical performance comparison among GUSS, GUS-nS, ZFS, SWF, SUS and exhaustive search. The achieved sum rate $R(S)$ and the number of selected users $|S|$ of those algorithms under different $K$ and $P$, averaged over channel distribution , are compared in the following.

### *A. Number of users*

The simulated multi-user system has $M = 10$ transmit antennas at BS, transmit SNR $P = 15\,dB$, and the number of users $K$ ranges from 8 to 20. All curves are obtained by averaging over $10^4$ independent complex-valued channels, drawn from i.i.d. Rayleigh distribution with unit-variance for each channel entry.

Fig. 5 shows that the throughput of all algorithms grows with the number of selected users. The reason encompasses two parts: first, the larger $K$ provides the higher multiuser diversity gain as there is more likeliness to select a user set with strong channel norm $|h|$ and effective-channel-gain $\lambda$; second, the larger $K$ provides the higher multiplexing gain because the cardinality of selected user set increases with $K$ as shown in Fig. 6.

The exhaustive search achieves the highest throughput of all user selection algorithms, which is followed sequentially by GUSS, GUS-nS, ZFS and SUS. The SUS is simulated with carefully chosen threshold $\alpha = 0.44$, which is optimum choice for $K = 13$, while the optimum $\alpha$ ranges between 0.41 and 0.52 when $K$ changes from 20 to 8. ZFS achieves considerable higher sum rate than SUS as it guarantees sum rate increment in each step of user selection.

To reveal more details on the performance of GUSS and GUS-nS algorithm, the ratio of eliminating redundant user and escaping from local optimum of these two algorithms, which corresponds to the ratio of user selection instant with effective 'delete' and 'swap' operation that increases sum rate, is presented in Fig. 7. GUS-nS achieves higher throughput than ZFS, 0.04 bps/Hz increment over ZFS for $K = 14$, by eliminating redundant users in $S_{ZFS}$ that the cardinality of selected user set $|S_{GUS-nS}| < |S_{ZFS}|$ as shown in Fig. 6. In average, $5.0\%$ of $S_{ZFS}$ contains redundant users according to Fig. 7.

GUSS achieves further throughput increment over ZFS, 0.43 bps/Hz increment over ZFS for $K = 14$, by eliminating redundant users and escaping from local optimum in $S_{ZFS}$ simultaneously. It selects a user set with larger cardinality, $|S_{GUSS}| > |S_{ZFS}|$, as shown in Fig. 6. It indicates that more effective 'add' operation with $\Delta R > 0$ is conducted after 'swap' operation, because only 'add' enlarges user set and 'swap' operation does not. According to Fig. 7, $40.1\%$ of $S_{ZFS}$ is trapped in local optimum in average and the ratio increases with $K$. The ratio of eliminating redundant user is $7.1\%$ in GUSS, which is higher than that in GUS-nS because the add operation after swap in GUSS will introduce more redundant users. GUSS achieves a higher sum rate and cardinality of user set than ZFS but still lower than exhaustive search as only one-for-one swap is used in GUSS.

### *B. Transmit SNR*

The achieved throughput and the cardinality of selected user set are both increased with the transmit SNR $P$, with the same trend as with $K$ in Fig. 5 and Fig. 6, for all algorithms except SUS. The SUS algorithm selects the same user set under different $P$ because its user selection procedure does not take $P$ into consideration. However, SUS achieves higher sum rate at larger $P$ as the sum rate increases with $P$ for the same user set.

Fig. 8 shows the throughputs of GUSS, GUS-nS, SWF and ZFS algorithms as a fraction of the throughput of exhaustive search algorithm at different transmit SNRs $P$. Fig. 9 shows the ratio of channel instants that has redundant user and local optimum encountered in the user selection process of GUSS, GUS-nS and SWF algorithms. The simulated multi-user system has $M = 10$, $K = 15$ and $P$ ranges from $0\,dB$ to $30\,dB$. All curves are obtained by averaging over $10^6$ independent channels.

The throughput ratios rank from high to low sequentially are GUSS, GUS-nS, and SWF and ZFS. The fraction of the GUSS throughput to the throughput of exhaustive search approaches 1 when $P$ approaches zero or infinity, and it exhibits a valley in the middle. The same trend exists for GUS-nS, SWF and ZFS but it requires higher $P$ for those algorithms to recover from the valley.

SWF has exactly the same sum rate performance with ZFS and the ratio of 'eliminating redundant user' for SWF equals to zero for the whole range considered in Fig. 9. There is no redundant user with $p_i = 0$ ever happened in one million simulations, which proofs the conclusion in Section III. GUS-nS achieves $98.2\%$ of sum rate upper bound in average, which corresponds to $0.1\%$ throughput increment over ZFS, by eliminating $4.2\%$ redundant users in $S_{ZFS}$ in average as shown in Fig. 9. The ratio of redundant user increases with $P$ from $0\,dB$ to $15\,dB$ and then decreases, because the redundant user existed when $P$ is low will not be redundant user any more when $P$ becomes large enough. Such as the example in Fig. 2, user 2 is a redundant user when $P = 30\,dB$ but is not when $P$ increases to $40\,dB$.

GUSS achieves $1.7\%$ higher sum rate than ZFS at $P = 30\,dB$ since there is at least $63.8\%$ of $S_{ZFS}$ trapped in local optimum and $5.2\%$ of $S_{ZFS}$ contains redundant user and they are all handled by GUSS as shown in Fig. 9. The gap between GUSS and ZFS increases with $P$ in the range shown in Fig. 8 because the possibility of the $S_{ZFS}$ trapped in local optimum increases with $P$. At the same time, GUSS eliminates $2.7\%$ more redundant user than GUS-nS in average because more effective 'add' operation with $\Delta R > 0$ is conducted after the 'swap' operation in GUSS, which turns more users to redundant user. In average, there is $6.9\%$ channel instants involve redundant user and $43.1\%$ channel instants are trapped local optimum in the process of GUSS. According to Fig. 8, GUSS achieves $99.3\%$ of sum rate upper bound averaged over the SNR range considered.

### C. Complexity of GUSS

GUSS provides considerable throughput increment over ZFS by adding the ‘delete’ and ‘swap’ operations which introduce a $2a+1$ linear complexity raise. The number of swap operation $a$ is influenced by $K$, $M$, $P$ and $H$. Fig. 10 shows the averaged $a$ for different number of users $K$ ranging from 10 to 40 at $P = 15\,dB$ and $M = 5, 10$. Fig. 11 shows the averaged $a$ for $10 \leq K \leq 40$, $0\,dB \leq P \leq 30\,dB$ at $M = 10$. All curves are obtained by averaging over $10^4$ independent channels.

For all $M$ and $K$ considered in Fig. 10, $a$ stays between 1.4 and 1.85 which implies that GUSS has four to five times complexity of ZFS. GUSS has more swap operations at $M = 10$ than at $M = 5$ for each specific $K$ when $K > M$. The fact that system with larger $M$ selects more users implies that the larger possibility $S_{ZFS}$ been trapped in local optimum. The $a$ decreases with $K$ when $K > 30$ for $M = 10$, and $K > 25$ for $M = 5$. Because the selected users are almost orthogonal with high probability when $K$ is large enough, it requires smaller $K$ to achieve near-orthogonal user set for smaller $M$ antennas in BS.

The $a$ stays between 1 and 2.5 for the $K$ and $P$ range considered in Fig. 11, which implies that GUSS has only three to six times complexity of ZFS. The $a$ increases with $K$ before saturated for given $P$, and it needs smaller $K$ to achieve the maximum $a$ at larger $P$. $a$ also increases with $P$ before saturated and then decrease with $P$, because the number of selected users increases with $P$ and saturated when $P$ is large enough. The $a$ equals to 2.04 at $P = 30\,dB$, $K = 15$ and $M = 10$, which corresponds to about five times complexity of ZFS for GUSS.

## VII. Conclusion

We have discovered two flaws in traditional greedy user selection in multi-user MIMO downlink with ZFBF: ‘redundant user’ and ‘local optimum’. While traditional greedy user selection methods only use ‘add’ operation during the update of the selected user set, the proposed GUSS algorithm allows ‘delete’ and ‘swap’ operations to eliminate redundant users and helps escaping from the local optimums. An ECV based effective-channel-gain $\lambda$ updating method for ‘add’, ‘delete’ and ‘swap’ user operation is designed to reduce the complexity of GUSS. The GUSS provides considerable throughput increment with only $2a+1$ linear complexity increase, where $a$ is the number of swap operations for specific realization and it stays between 1 and 2.5 according to our simulation results. Simulation results verify the improved throughput performance and low complexity.

The GUSS algorithm proposed in this paper achieves $99.3\%$ of the upper bound throughput performance; it is significant for multi-user MIMO downlink transmission. And the novel ECV based efficient channel gain $\lambda$ updating method is a useful component to build more delicate user selection algorithms, such as the decremental user selection algorithm proposed for massive multi-antenna system in [20]. The work in this paper can be extended in several ways, including considering per-antenna transmit power constraint, multi-antenna users, partial CSIT, and user fairness among users.

## APPENDIX

Proof of Lemma 1: Suppose the selected user set with redundant user is $S = \{1, 2, \cdots, n\}$, the ECV and effective-channel-gain of the user $i \in S$ is $\boldsymbol{\nu}_i$ and $\lambda_i$, respectively. Let $\lambda_1 \geq \lambda_2 \geq \cdots \geq \lambda_n$ and only the user $n$ is allocated with zero transmit power that

$$\frac{1}{\lambda_{n-1}} \leq \mu \leq \frac{1}{\lambda_n}.$$

where $\mu = \frac{1}{n-1}\left(P + \sum_{i=1}^{n-1} \frac{1}{\lambda_i}\right)$ is the water level for $S$. Suppose deleting user $k$ achieves the maximum sum rate among $S \setminus \{j\}$, i.e., $k = \arg\max_{j \in S} R(S \setminus \{j\})$. The conclusion of Lemma 1 equals to

$$R(S \setminus \{n\}) \geq R(S) \tag{61}$$

and

$$R(S \setminus \{k\}) > R(S \setminus \{n\}). \tag{62}$$

Denote the updated effective-channel-gain of user $i$ after deleting user $j \in S$ as $\lambda_{i,j-}$ and the corresponding water level as $\mu_{j-}$, according to (48) we have

$$\lambda_{i,j-} = \frac{\lambda_i^2 \lambda_j}{\lambda_i \lambda_j - \left\|\boldsymbol{\nu}_i \boldsymbol{\nu}_j^*\right\|^2}, \quad i \in S \setminus \{j\}$$

and $\lambda_{i,j-} \geq \lambda_i$ for all $i \in S \setminus \{j\}$ since $\lambda_i \lambda_j \geq \|\boldsymbol{\nu}_i \boldsymbol{\nu}_j^*\|^2$. According to (10), the (61) holds

because

$$
\begin{aligned}
R(S \setminus \{n\}) &\geq \sum_{i \neq n} \log \left( 1 + (\mu - \frac{1}{\lambda_i}) \lambda_{i,n-} \right) \\
&\geq \sum_{i \neq n} \log \left( 1 + (\mu - \frac{1}{\lambda_i}) \lambda_i \right) \\
&= R(S)\,,
\end{aligned} \tag{63}
$$

where the first inequality holds as $S \setminus \{n\}$ achieves equal or larger sum rate than distributing power the same as that in $S$, and the second inequality holds since $\lambda_{i,n-} \geq \lambda_i$.

Suppose the transmit power scaling factor of user $i$ in $S \setminus \{j\}$ is $p_{i,j-}$ after waterfilling. The (62) holds on the condition

$$
\prod_{i \neq k, p_{i,k-} > 0} \frac{\mu_{k-} \lambda_i}{\sin^2 \theta_{i,k}} > \prod_{i \neq n, p_{i,n-} > 0} \frac{\mu_{n-} \lambda_i}{\sin^2 \theta_{i,n}}\,. \tag{64}
$$

where $\theta_{i,j}$ is the angle between $\boldsymbol{\nu}_i$ and $\boldsymbol{\nu}_j$ that is independent of $\lambda_i$ and $\lambda_j$, $\cos^2 \theta_{i,\,j} = \frac{\left\| \boldsymbol{\nu}_i \boldsymbol{\nu}_j^* \right\|^2}{\lambda_i \lambda_j}$. The (64) is achievable when the user $k$ has stronger channel correlation with the other users than that of the user $n$, i.e., $\sin^2 \theta_{i,k} < \sin^2 \theta_{i,n}$ and deleting user $k$ provides larger ECV increment for user $i \in S \setminus \{k, n\}$ that $\lambda_{i,k-} > \lambda_{i,n-}$. The throughput increment in users $i \in S \setminus \{k, n\}$ could compensate the throughput loss in deleing the user $k$.

## References

[1] D. Gesbert, M. Kountouris, R. W. Heath, C. B. Chae, and T. Salzer, "Shifting the mimo paradigm," *IEEE Signal Process. Mag.*, vol. 24, pp. 36–46, Sep. 2007.

[2] G. Dimic and N. D. Sidiropoulos, "On downlink beamforming with greedy user selection: Performance analysis and a simple new algorithm," *IEEE Trans. Signal Process.*, vol. 53, pp. 3857–3868, Oct. 2005.

[3] T. Yoo and A. Goldsmith, "On the optimality of multiantenna broadcast scheduling using zero-forcing beamforming," *IEEE J. Sel. Areas Commun.*, vol. 24, pp. 528–541, Mar. 2006.

[4] Z. K. Shen, R. H. Chen, J. G. Andrews, R. W. Health, and B. L. Evans, "Low complexity user selection algorithms for multiuser mimo systems with block diagonalization," *IEEE Trans. Signal Process.*, vol. 54, pp. 3658–3663, Sep. 2006.

[5] J. Q. Wang, D. J. Love, and M. D. Zoltowski, "User selection with zero-forcing beamforming achieves the asymptotically optimal sum rate," *IEEE Trans. Signal Process.*, vol. 56, pp. 3713–3726, Aug. 2008.

[6] M. Costa, "Writing on dirty paper," *IEEE Trans. Inf. Theory*, vol. 29, pp. 439–441, May 1983.

[7] G. Caire and S. Shamai, "On the achievable throughput of a multiantenna gaussian broadcast channel," *IEEE Trans. Inf. Theory*, vol. 49, pp. 1691–1706, Jul. 2003.

[8] S. Vishwanath, N. Jindal, and A. G. Goldsmith, "Duality, achievable rates, and sum-rate capacity of gaussian mimo broadcast channels," *IEEE Trans. Inf. Theory*, vol. 49, pp. 2658–2668, Oct. 2003.

[9] W. Yu and J. M. Cioffi, "Sum capacity of gaussian vector broadcast channels," *IEEE Trans. Inf. Theory*, vol. 50, pp. 1875–1892, Sep. 2004.

[10] P. Viswanath and D. N. C. Tse, "Sum capacity of the vector gaussian broadcast channel and uplink-downlink duality," *IEEE Trans. Inf. Theory*, vol. 49, pp. 1912–1921, Aug. 2003.

[11] L. U. Choi and R. D. Murch, "A transmit preprocessing technique for multiuser mimo systems using a decomposition approach," *IEEE Trans. Wireless Commun.*, vol. 3, pp. 20–24, Jan. 2004.

[12] A. Wiesel, Y. C. Eldar, and S. Shamai, "Zero-forcing precoding and generalized inverses," *IEEE Trans. Signal Process.*, vol. 56, pp. 4409–4418, Sep. 2008.

[13] Q. H. Spencer, A. L. Swindlehurst, and M. Haardt, "Zero-forcing methods for downlink spatial multiplexing in multiuser mimo channels," *IEEE Trans. Signal Process.*, vol. 52, pp. 461–471, Feb. 2004.

[14] C. B. Peel, B. M. Hochwald, and A. L. Swindlehurst, "A vector-perturbation technique for near-capacity multiantenna multiuser communication - part i: Channel inversion and regularization," *IEEE Trans. Commun.*, vol. 53, pp. 195–202, Jan. 2005.

[15] T. Haustein, C. von Helmolt, E. Jorswieck, V. Jungnickel, and V. Pohl, "Performance of mimo systems with channel inversion," in *55st IEEE Vehicular Technology Conference*, May 2002, pp. 35–39.

[16] S. Akhlaghi, A. K. Khandani, and A. Falahati, "User selection and signaling over time-varying mimo broadcast channels," in *23rd Biennial Symposium on Communications*, May 2006, pp. 31–34.

[17] L. Jin, X. Gu, and Z. Hu, "Low-complexity scheduling strategy for wireless multiuser multiple-input multiple-output downlink system," *IET Commun.*, vol. 5, pp. 990–995, May 2011.

[18] C. D. Meyer, *Matrix analysis and applied linear algebra*. SIAM: Society for Industrial and Applied Mathematics, 2001.

[19] D. Tse and P. Viswanath, *Fundamentals of Wireless Communication*. Cambridge University Press, 2005.

[20] S. Huang, H. Yin, J. Wu, and V. C. M. Leung, "Decremental user selection for massive multi-user mimo downlink with zero-forcing beamforming," *Submitted to IEEE Wireless Commun. Lett.*

TABLE I: Comparison between ZFS user selection and exhaustive search

| Transmit SNR | Procedure of ZFS | $S_{ZFS}$ | $S^{best}$ |
|---|---|---|---|
| $0 \leq P \leq 10.48$ | $S_1 = \{2\}$ | $\{2\}$ | $\{1,3\}$ |
| $10.48 < P \leq 27.13$ | $S_1 = \{2\}, S_2 = \{1, 2\}$ | $\{1, 2\}$ | $\{1, 3\}$ |
| $27.13 < P \leq 34.85$ | $S_1 = \{2\}, S_2 = \{1, 2\}, S_3 = \{1, 2, 3\}$ | $\{1, 2, 3\}$ | $\{1, 3\}$ |
| $P > 34.85$ | $S_1 = \{2\}, S_2 = \{1, 2\}, S_3 = \{1, 2, 3\}$ | $\{1, 2, 3\}$ | $\{1, 2, 3\}$ |

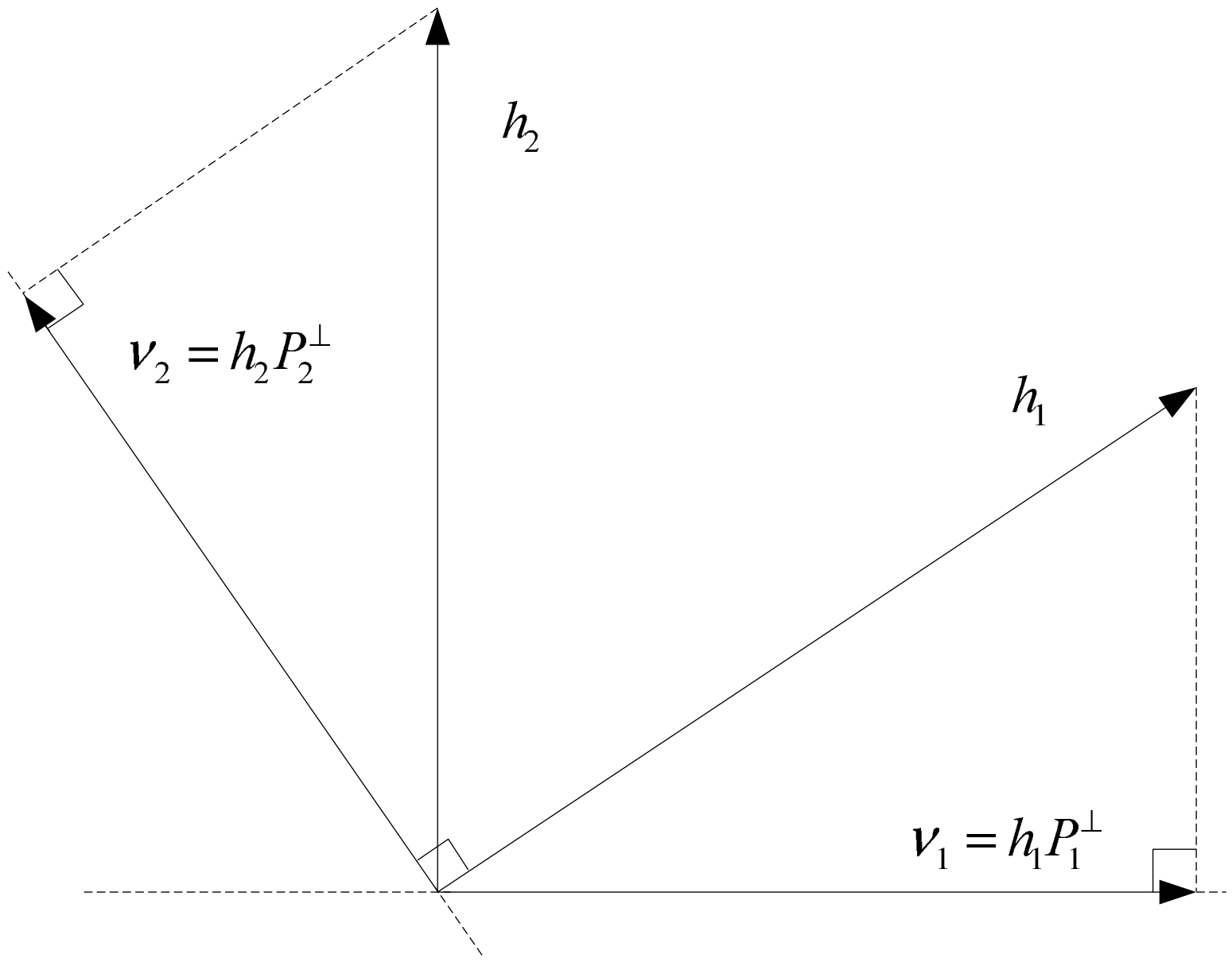

Fig. 1: An example of ECV calculation when selected user set $S = \{1, 2\}$

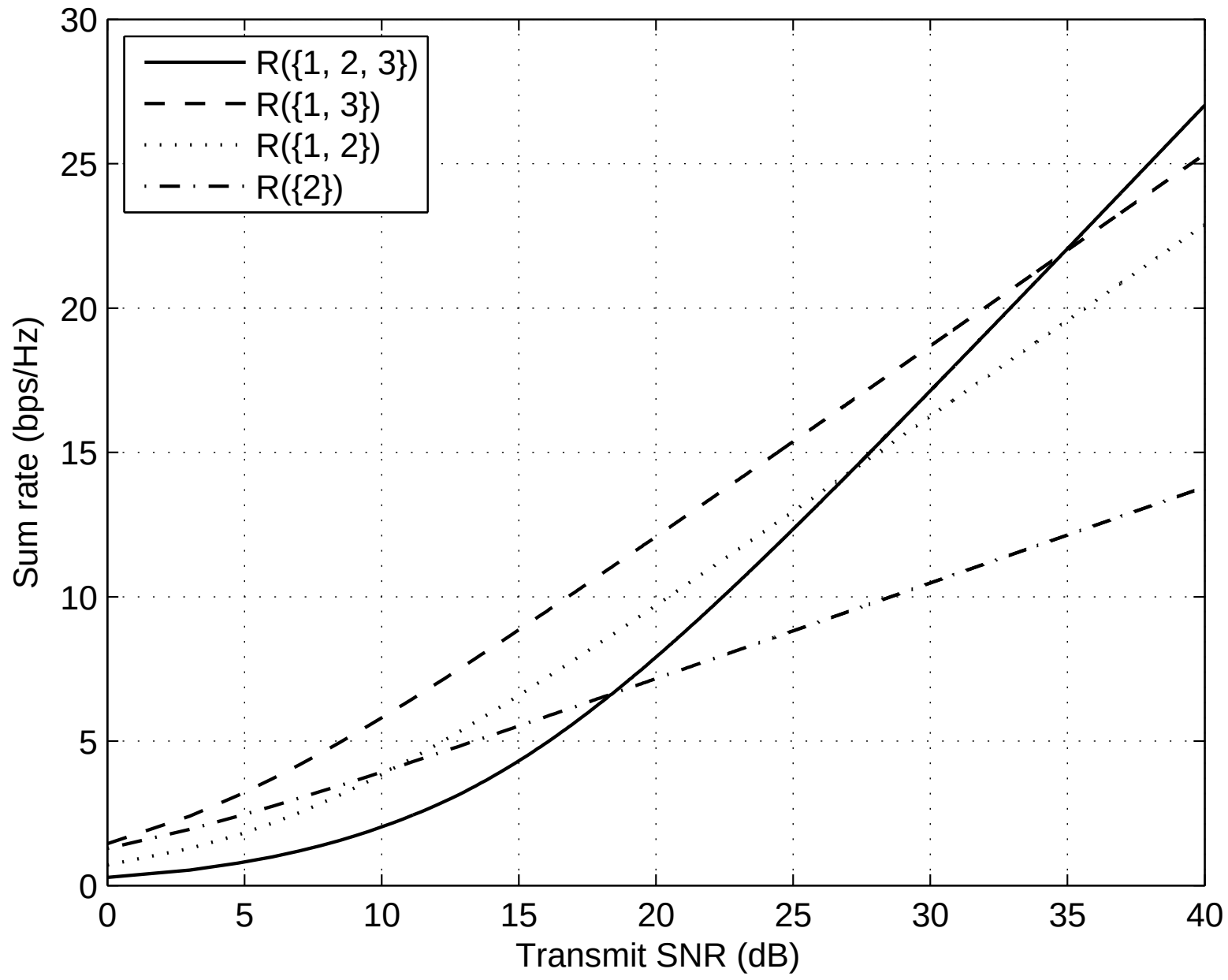

Fig. 2: Sum rate versus transmit SNR for different selected user sets

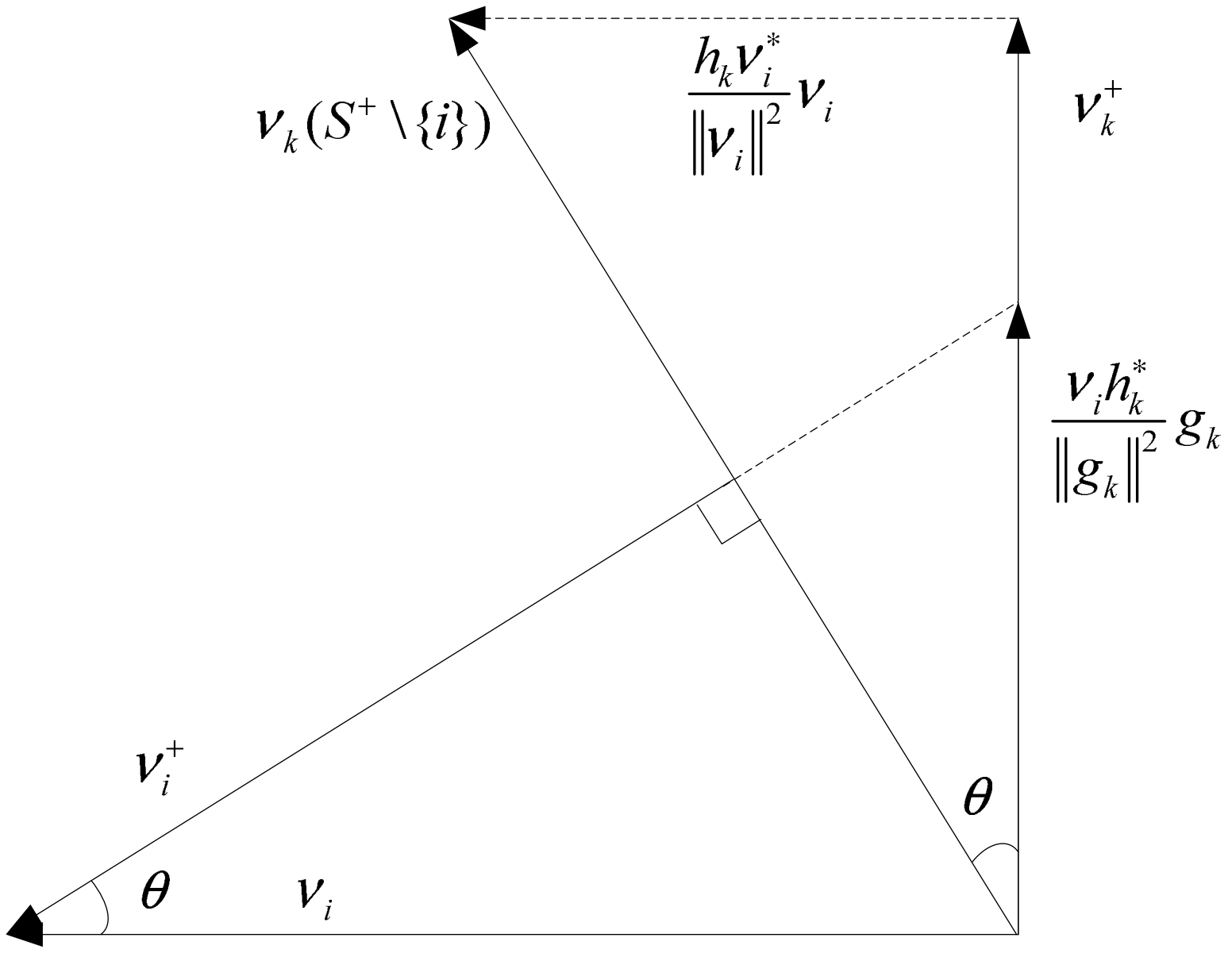

Fig. 3: ECV update for user $i$ after adding a new user $k$

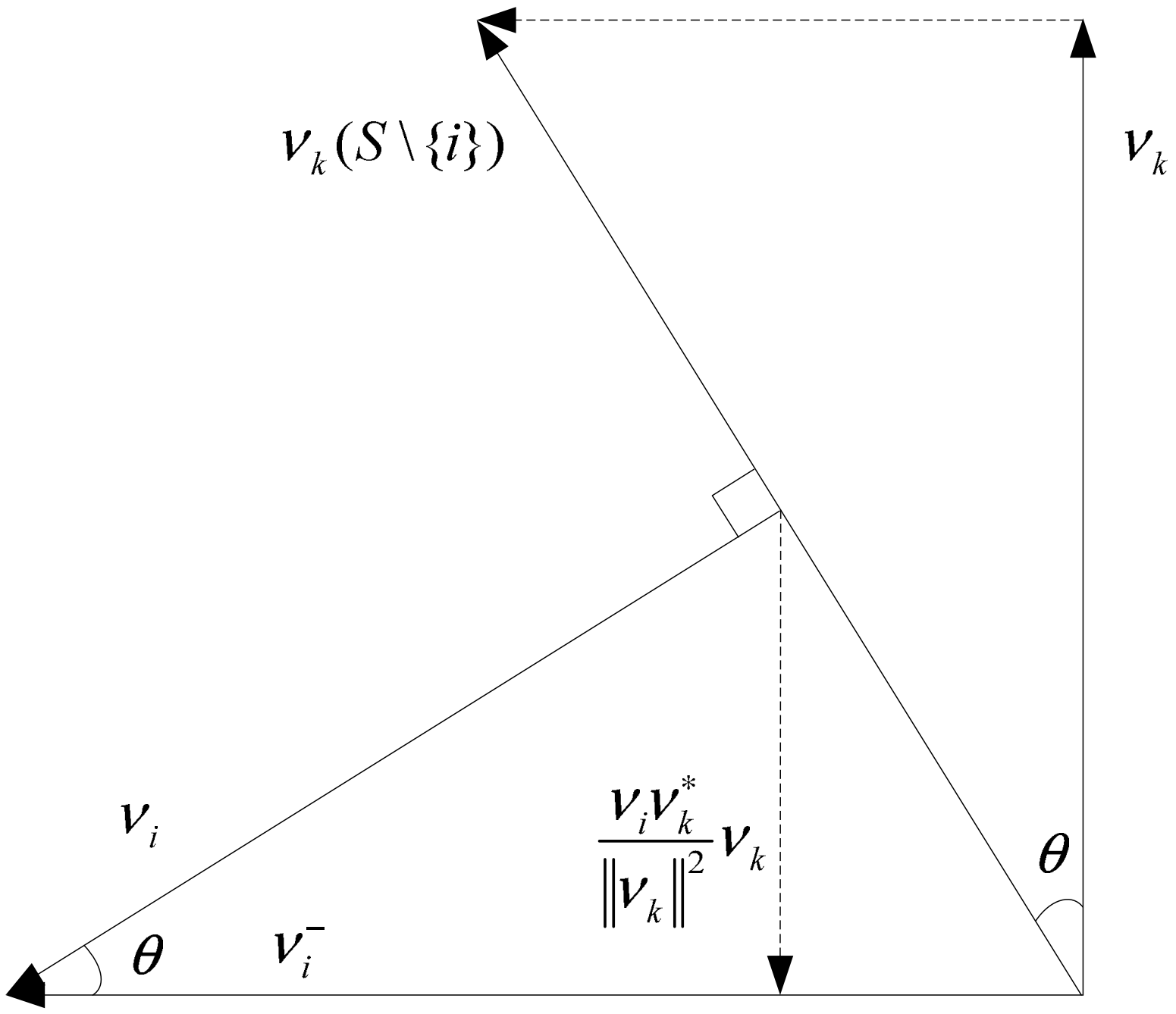

Fig. 4: ECV update for user $i$ after deleting a selected user $k$

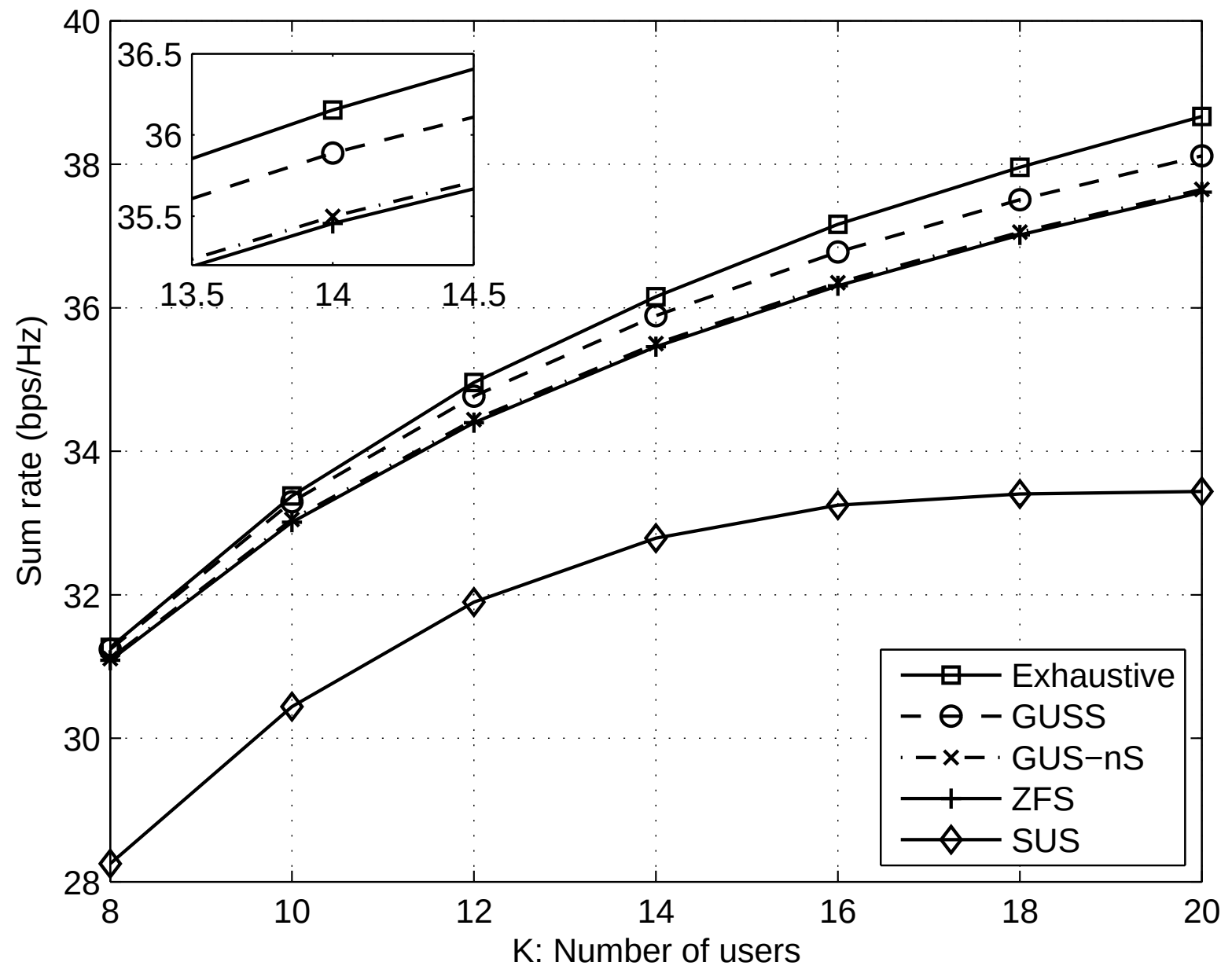

Fig. 5: Sum rate performance comparison of GUSS, GUS-nS, ZFS, SUS and exhaustive search algorithms with $M = 10$ and $P = 15\,dB$

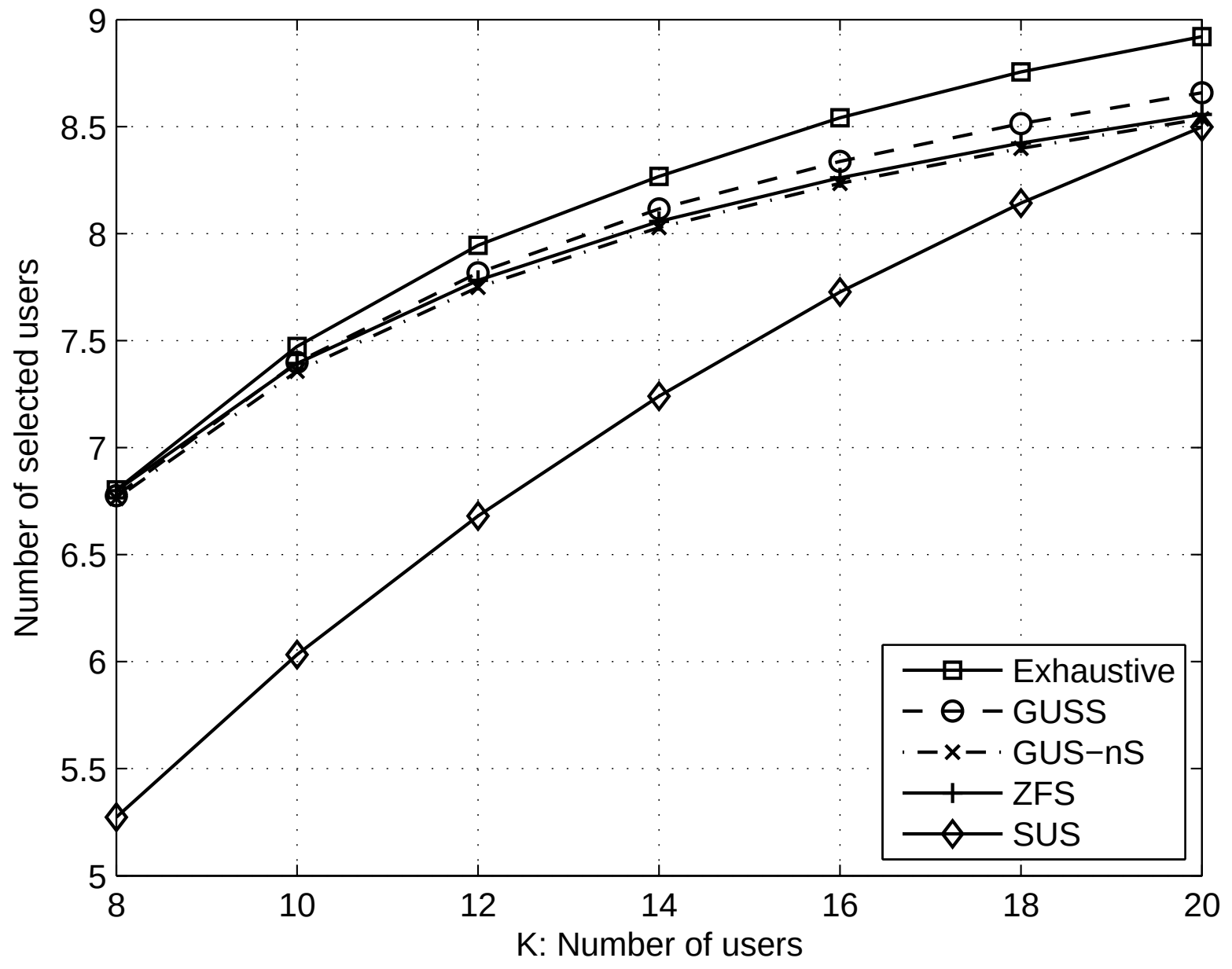

Fig. 6: Cardinality of selected user set comparison of GUSS, GUS-nS, ZFS, SUS and exhaustive search algorithms with $M = 10$ and $P = 15\,dB$

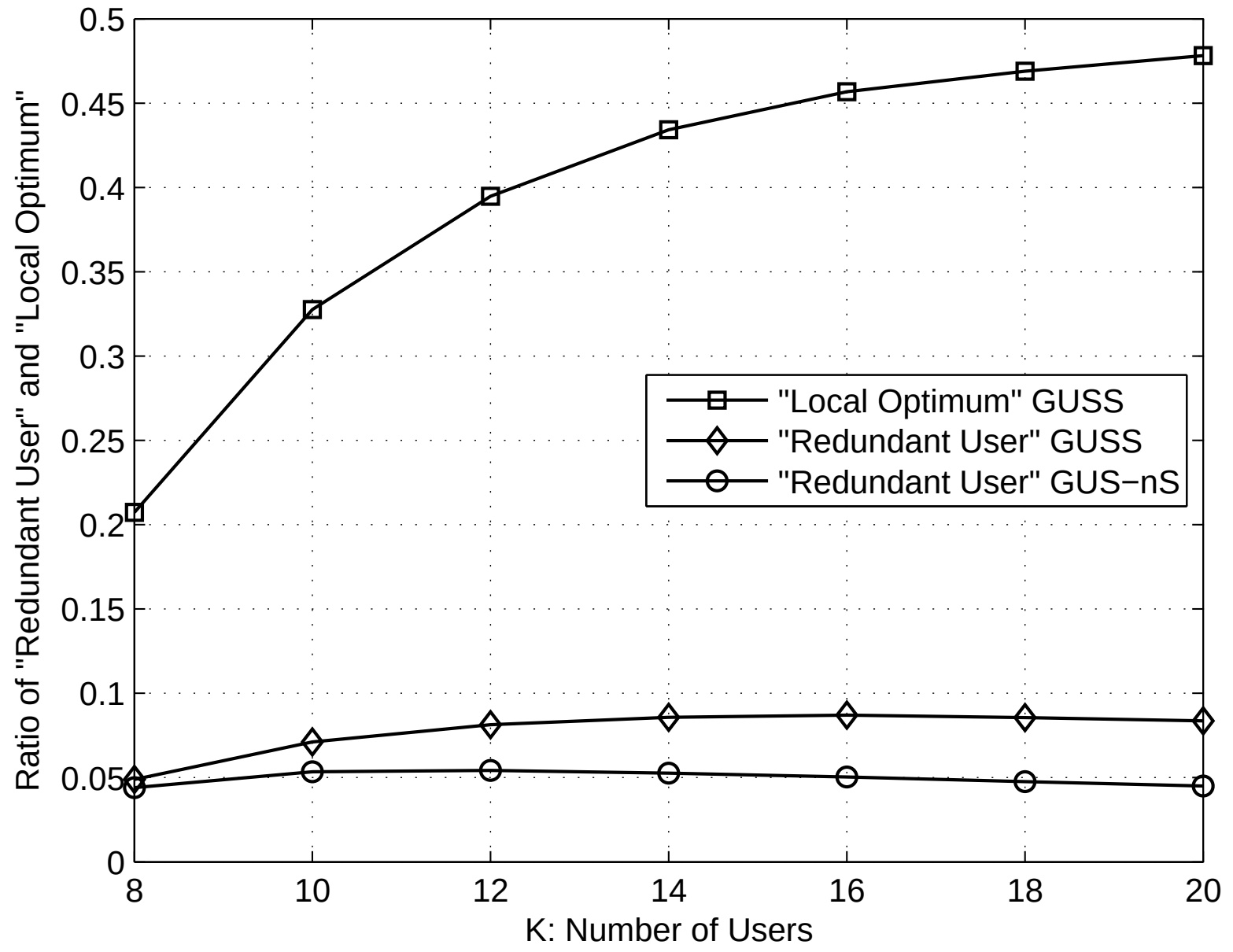

Fig. 7: Ratio of GUSS and GUS-nS algorithms ‘eliminate redundant user’ and ‘escape from local optimum’ with $M = 10$ and $P = 15\,dB$

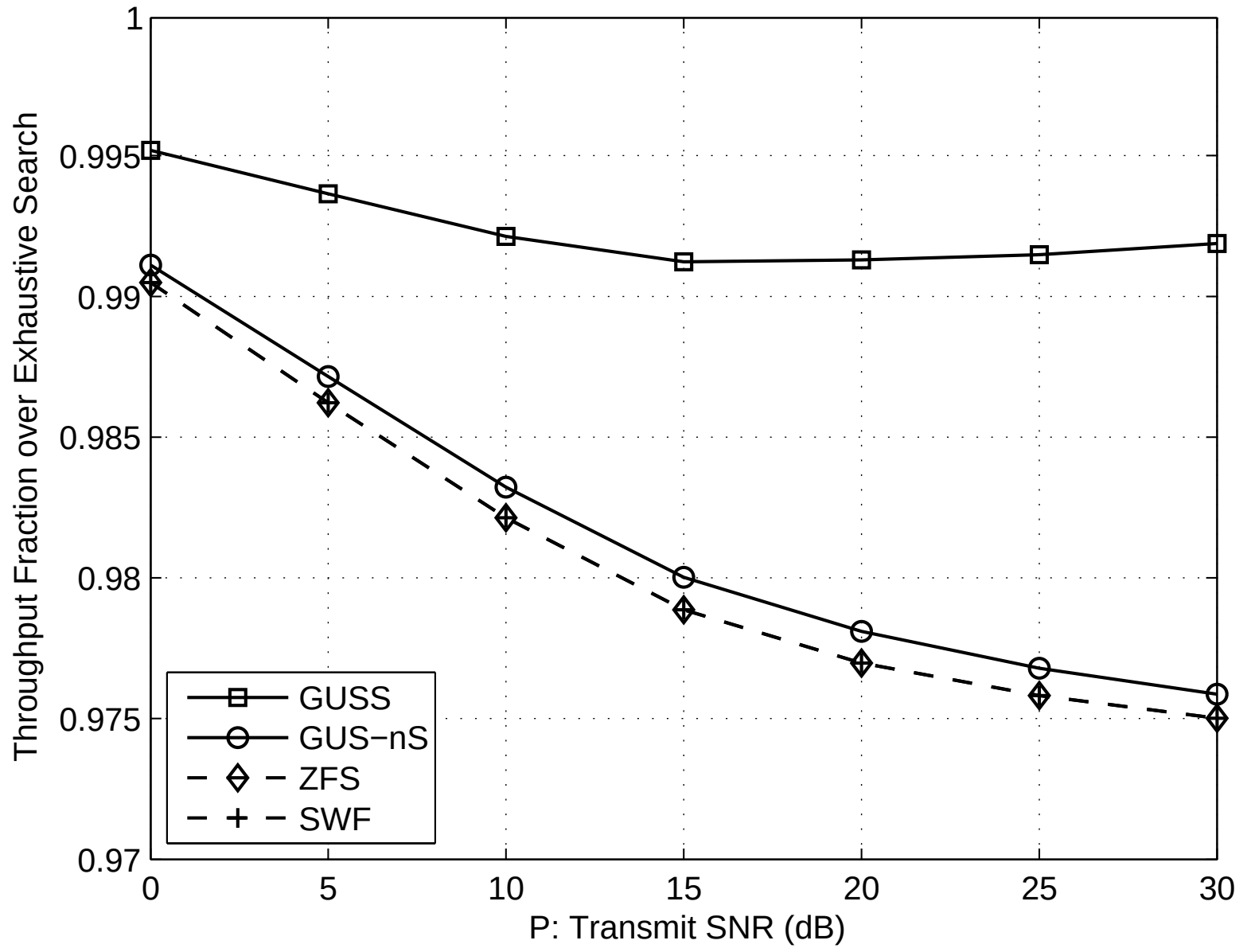

Fig. 8: Throughput fractions of GUSS, GUS-nS, ZFS and SWF algorithms over the throughput of exhaustive search with $M = 10$ and $K = 15$

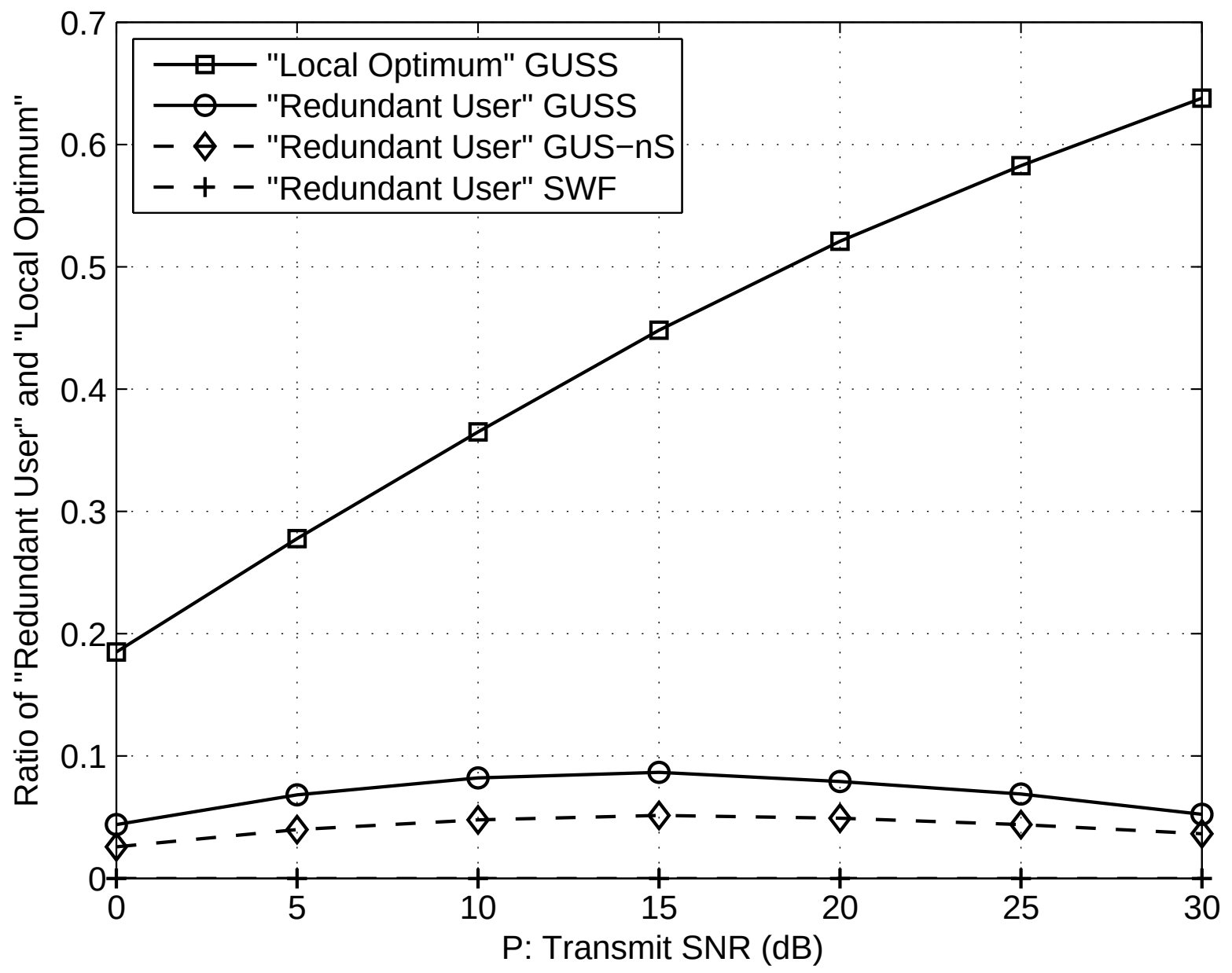

Fig. 9: Ratio of GUSS, GUS-nS and SWF algorithm ‘eliminating redundant user’ and ‘escaping from local optimum’ with $M = 10$ and $K = 15$

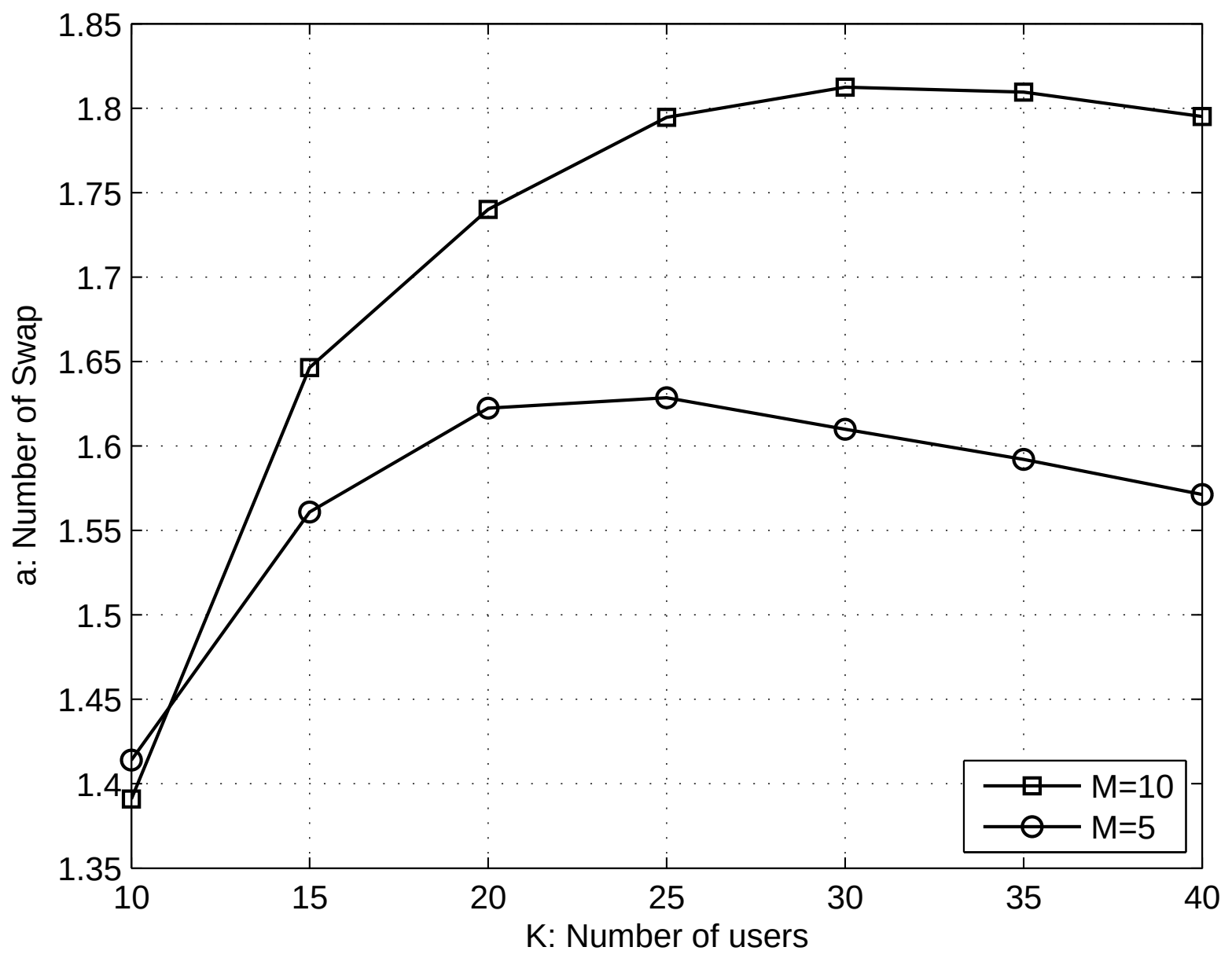

Fig. 10: Number of swaps in GUSS for different number of users $K$ at $P = 15\,dB$ and $M = 5, 10$.

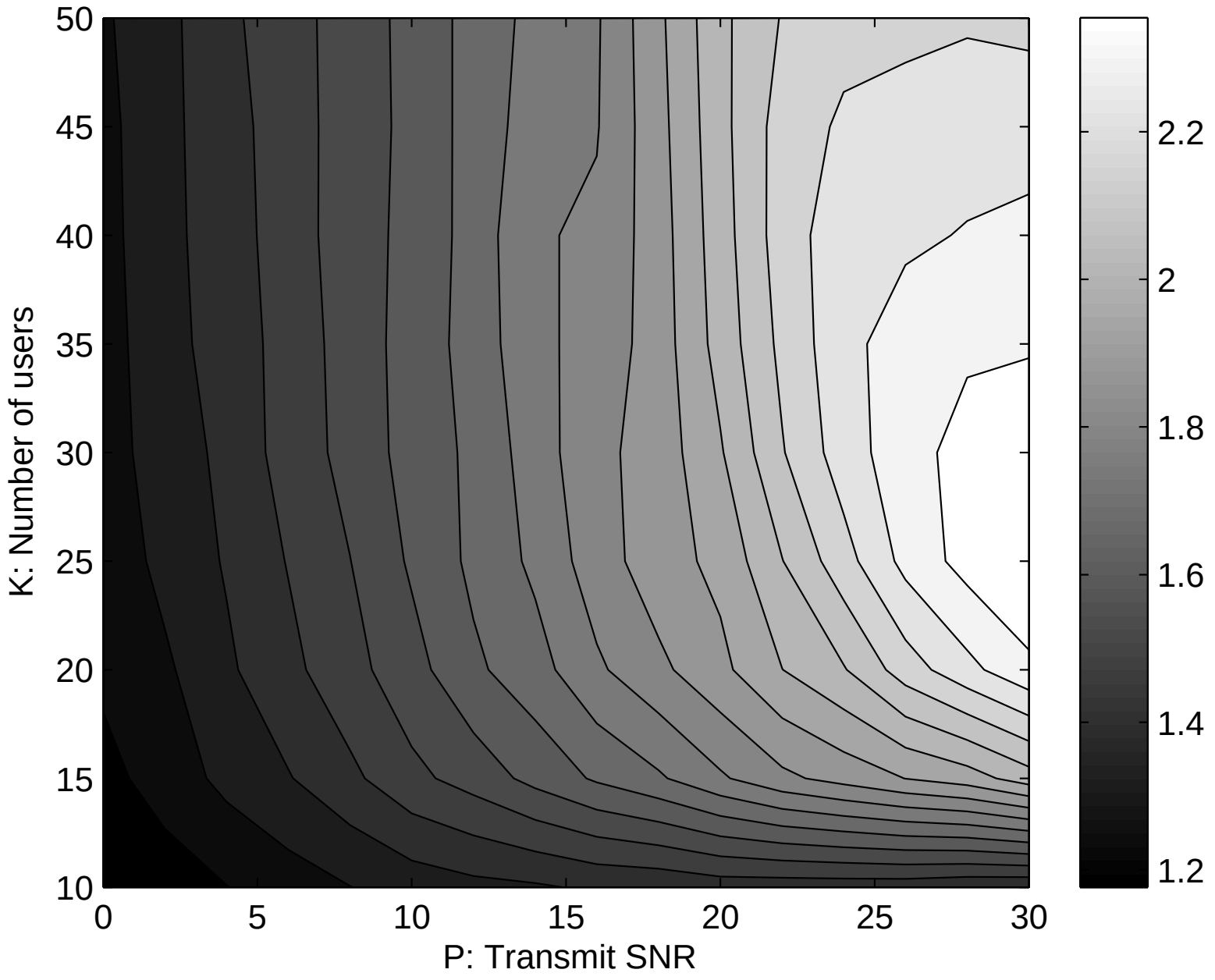

Fig. 11: Number of swaps $a$ in GUSS for $10 \leq K \leq 50$ and $0\,dB \leq K \leq 30\,dB$ at $M = 10$